\let\includefigures=\iftrue
%
\let\useblackboard=\iftrue
%
%
\newfam\black
\input harvmac.tex
\input epsf
\includefigures
\message{If you do not have epsf.tex (to include figures),}
\message{change the option at the top of the tex file.}
\def\figin{\epsfcheck\figin}\def\figins{\epsfcheck\figins}
\def\epsfcheck{\ifx\epsfbox\UnDeFiNeD
\message{(NO epsf.tex, FIGURES WILL BE IGNORED)}
\gdef\figin##1{\vskip2in}\gdef\figins##1{\hskip.5in}
\else\message{(FIGURES WILL BE INCLUDED)}%
\gdef\figin##1{##1}\gdef\figins##1{##1}\fi}
\def\DefWarn#1{}
\def\figinsert{\goodbreak\midinsert}
\def\ifig#1#2#3{\DefWarn#1\xdef#1{fig.~\the\figno}
\writedef{#1\leftbracket fig.\noexpand~\the\figno}%
\figinsert\figin{\centerline{#3}}\medskip\centerline{\vbox{\baselineskip12pt
\advance\hsize by -1truein\noindent\footnotefont{\bf Fig.~\the\figno:} #2}}
\bigskip\endinsert\global\advance\figno by1}
\else
\def\ifig#1#2#3{\xdef#1{fig.~\the\figno}
\writedef{#1\leftbracket fig.\noexpand~\the\figno}%
\global\advance\figno by1}
\fi
\useblackboard
\message{If you do not have msbm (blackboard bold) fonts,}
\message{change the option at the top of the tex file.}
\font\blackboard=msbm10 scaled \magstep1
\font\blackboards=msbm7
\font\blackboardss=msbm5
\textfont\black=\blackboard
\scriptfont\black=\blackboards
\scriptscriptfont\black=\blackboardss

\else

\fi
\def\yboxit#1#2{\vbox{\hrule height #1 \hbox{\vrule width #1
\vbox{#2}\vrule width #1 }\hrule height #1 }}
\def\fillbox#1{\hbox to #1{\vbox to #1{\vfil}\hfil}}
\def\ybox{{\lower 1.3pt \yboxit{0.4pt}{\fillbox{8pt}}\hskip-0.2pt}}
\def\mapr{\mathop{\longrightarrow}\limits}
\def\grade{\varphi}

\def\comments#1{}

\def\p{\partial}

\def\half{{1\over 2}}

\def\tr{{\rm tr\ }}

\def\Im{{\rm Im\hskip0.1em}}
\def\Ext{{\rm Ext}}

\def\CE{{\cal E}}

\def\CM{{\cal M}}
\def\CN{{\cal N}}
\def\CO{{\cal O}}

\def\CQ{{\cal Q}}

\def\sgn{{\rm sgn\ }}
\def\P{\BP}
\def\I{I}

\def\II{\relax{I\kern-.10em I}}
\def\IIa{{\II}a}

\def\IZ{\relax\ifmmode\mathchoice
{\hbox{\cmss Z\kern-.4em Z}}{\hbox{\cmss Z\kern-.4em Z}}
{\lower.9pt\hbox{\cmsss Z\kern-.4em Z}}
{\lower1.2pt\hbox{\cmsss Z\kern-.4em Z}}\else{\cmss Z\kern-.4em
Z}\fi}
\def\IB{\relax{\rm I\kern-.18em B}}
\def\IC{{\relax\hbox{$\inbar\kern-.3em{\rm C}$}}}
\def\ID{\relax{\rm I\kern-.18em D}}
\def\IE{\relax{\rm I\kern-.18em E}}
\def\IF{\relax{\rm I\kern-.18em F}}
\def\IG{\relax\hbox{$\inbar\kern-.3em{\rm G}$}}
\def\IGa{\relax\hbox{${\rm I}\kern-.18em\Gamma$}}
\def\IH{\relax{\rm I\kern-.18em H}}
\def\II{\relax{\rm I\kern-.18em I}}
\def\IK{\relax{\rm I\kern-.18em K}}
\def\IP{\relax{\rm I\kern-.18em P}}

%
\def\inbar{\,\vrule height1.5ex width.4pt depth0pt}

\def\p{\partial}

\font\cmss=cmss10 \font\cmsss=cmss10 at 7pt
\def\IR{\relax{\rm I\kern-.18em R}}

\def\im{{\rm im\ }}
\def\ker{{\rm ker\ }}

\def\End{{\rm End\ }}
\def\Hom{{\rm Hom}}

\def\rk{{r}}
\def\chch{\hbox{ch}}
\def\chcl{\hbox{c}}

\def\aroof{\hat A}
\def\BR{\IR}
\def\BZ{Z} 
\def\BP{\IP}
\def\BR{\IR}
\def\BC{\IC}

\def\BN{N}

\def\lp10{l_P^{10}}
\def\lp11{l_P^{11}}
\def\R11{R_{11}}

\Title{\hbox{RU-2000-07}}
{\vbox{
\centerline{The spectrum of BPS branes on a noncompact Calabi-Yau}}}
\smallskip
\centerline{Michael R. Douglas\footnote{$^{\&}$}{
Louis Michel Professor}, Bartomeu Fiol and Christian R\"omelsberger}
\medskip
\centerline{Department of Physics and Astronomy}
\centerline{Rutgers University }
\centerline{Piscataway, NJ 08855--0849}
\medskip
\centerline{$^{\&}$ I.H.E.S., Le Bois-Marie, Bures-sur-Yvette, 91440 France}
\medskip
\centerline{\tt mrd, fiol, roemel @physics.rutgers.edu}
\bigskip
\noindent
We begin the study of the spectrum of BPS branes and its
variation on lines of marginal stability on $\CO_{\P^2}(-3)$, 
a Calabi-Yau ALE space asymptotic to $\BC^3/\BZ_3$.
We show how to get the complete spectrum near the large volume limit and near
the orbifold point, and find a striking similarity between the
descriptions of holomorphic bundles and BPS branes in these two limits.
We use these results to develop a general picture of the spectrum.
We also suggest a generalization of some of the ideas to the quintic
Calabi-Yau.

\Date{March 2000}
\nref\abpss{C. Angelantonj, M. Bianchi, G. Pradisi, A. Sagnotti and Y. Stanev,
``Chiral Asymmetry in Four-Dimensional Open-String Vacua,''
Phys. Lett. B385 (1996) 96-102, hep-th/9606169.}
\nref\agm{P. Aspinwall, B. R. Greene and D. R. Morrison,
Nucl. Phys. B420 (1994) 184-242; hep-th/9311042.}
\nref\bachas{C.~P.~Bachas, P.~Bain and M.~B.~Green, ``Curvature terms 
in D-brane actions and their M-theory origin,'' JHEP {\bf 9905}, 011 (1999)
hep-th/9903210.}
\nref\hulekb{W.~Barth and K.~Hulek, ``Monads and moduli of vector bundles'', 
Manuscripta Math. 25 (1978), no. 4, 323--347}
\nref\beilinson{A.~A.~Beilinson, ``Coherent Sheaves on $\P^n$ and Problems 
of Linear Algebra'', Funct. Anal. Appl. 12 (1978), 214-216.}
\nref\benson{D.J. Benson, {\it Representations and cohomology}, Cambridge 
University Press, 1991.}
\nref\brunscho{I. Brunner and V. Schomerus, 
``D-branes at Singular Curves of Calabi-Yau Compactifications,''
hep-th/0001132.}
\nref\brs{I. Brunner and V. Schomerus, to appear.}
\nref\bdlr{I.~Brunner, M.~R.~Douglas, A.~Lawrence and C.~Romelsberger,
``D-branes on the quintic,'' hep-th/9906200.}
\nref\calabi{E, Calabi, Ann. Sci. Ec. Norm. Sup. 12 (1979) 269-294.}
\nref\mirror{
T.-M. Chiang, A. Klemm, S.-T. Yau, and E. Zaslow,
``Local Mirror Symmetry: Calculations and Interpretations,''
hep-th/9903053.}
\nref\morezthree{M. Cveti\v c, L. Everett, P. Langacker, and J. Wang,
``Blowing-Up the Four-Dimensional $Z_3$ Orientifold,'' 
JHEP 9904 (1999) 020; hep-th/9903051.}
\nref\ddm{D.-E. Diaconescu, M. R. Douglas and D. R. Morrison, to appear.}
\nref\dg{D.-E.~Diaconescu and J.~Gomis, ``Fractional branes and boundary
states in orbifold theories,'' hep-th/9906242.}
\nref\diro{D.-E.~Diaconescu and C.~Romelsberger, ``D-branes and bundles 
on elliptic fibrations,'' hep-th/9910172.}
\nref\donaldson{S. K. Donaldson and P. B. Kronheimer, {\it The Geometry
of Four-Manifolds,} Oxford Univ. Press, 1990.}
\nref\dreview{M.R.~Douglas, ``Topics in D-geometry,'' hep-th/9910170.}
\nref\dtoappear{M.~R.~Douglas, ``D-branes and Categories,'' to appear.}
\nref\pistability{M.~R.~Douglas, B.~Fiol and C.~Romelsberger,
``Stability and BPS branes,'' hep-th/0002037.}
\nref\dgm{M.R.~Douglas, B.R.~Greene and D.R.~Morrison, ``Orbifold
resolution by D-branes,'' Nucl.\ Phys.\ {\bf B506}, 84 (1997)
hep-th/9704151.}
\nref\dm{M.R.~Douglas and G.~Moore, ``D-branes, Quivers, and ALE Instantons,''
hep-th/9603167.}
\nref\friedman{R. Friedman, {\it Algebraic Surfaces and Holomorphic
Vector Bundles}, Springer 1998.}
\nref\fmw{R. Friedman, J. Morgan and E. Witten,
``Vector Bundles over Elliptic Fibrations,'' alg-geom/9707029.}
\nref\gm{S.I. Gelfand and Y.I. Manin, {\it Methods of homological algebra}, 
Springer 1996}
\nref\gov{S. Govindarajan and T. Jayaraman,
``On the Landau-Ginzburg description of Boundary CFTs and special Lagrangian
submanifolds,'' hep-th/0003242.}
\nref\lazaroiu{B. R. Greene and C. I. Lazaroiu, hep-th/0001025;
C. I. Lazaroiu, hep-th/0002004.}
\nref\gh{P.~Griffiths, J.~Harris, {\it Principles of Algebraic Geometry}, 
John Wiley \& Sons, Inc. 1994}
\nref\hm{J. Harvey and G. Moore, ``On the algebras of BPS states'', 
 Commun.Math.Phys. 197 (1998) 489-519}
\nref\yang{Y.-H. He ``Some Remarks on the Finitude of Quiver Theories'', 
hep-th/9911114}
\nref\horja{R.~P.~Horja, ``Hypergeometric functions and mirror symmetry in toric varieties'', math.AG/9912109.}
\nref\huleka{K.~Hulek, ``On the Classification of Stable Rank-r Vector
Bunbles over the Projective Plane'', {\it Vector bundles and differential 
equations} (Proc. Conf., Nice, 1979), pp. 113--144, 
Progr. Math., 7, Birkhäuser, Boston, Mass., 1980. }
\nref\naka{Y. Ito and H. Nakajima,
``McKay correspondence and Hilbert schemes in dimension three,''
math.AG/9803120 }
\nref\kac{V.~Kac, ``Infinite Root Systems, Representations of
Graphs and Invariant Theory'', Invent. Math. 56 (1980) 57}
\nref\kapranov{M. M. Kapranov, ``On the derived categories of coherent 
sheaves on some homogeneous spaces'', Invent. Math. 92 (1988),no. 3, 479}
\nref\lerche{
P.Kaste, W.Lerche, C.A.Lutken, and J.Walcher, 
``D-Branes on K3-Fibrations,'' hep-th/9912147.}
\nref\katzklemmvafa{S. Katz, A. Klemm and C. Vafa, 
``M-Theory, Topological Strings and Spinning Black Holes,'' hep-th/9910181.}
\nref\king{A.~D.~King, ``Moduli of Representations of Finite Dimensional
Algebras'', Quart. J. Math. Oxford (2), 45 (1994), 515-530.}
\nref\stringnet{A. Klemm, W. Lerche, P. Mayr, C.Vafa and N. Warner,
``Self-Dual Strings and N=2 Supersymmetric Field Theory,''
Nucl.Phys. B477 (1996) 746-766, hep-th/9604034;
A. Sen and B. Zwiebach, 
``Stable Non-BPS States in F Theory,'' JHEP 003 (2000) 036,
hep-th/9907164;
T. Hauer and A. Iqbal, ``Del Pezzo Surfaces and Affine 7-Brane Backgrounds,''
JHEP 001 (2000) 043, hep-th/9910054;
K. Hori, A. Iqbal and C. Vafa, ``D-Branes and Mirror Symmetry,''
hep-th/0005247.}
\nref\kontsevich{M. Kontsevich,
 ``Homological algebra of mirror symmetry,''
alg-geom/9411018.}
\nref\kraft{H. Kraft and C. Riedtmann, 
``Geometry of representations of quivers,''
109-145. London Math. Soc. Lecture Note Ser., 116, Cambridge Univ. Press, 
1986.}
\nref\lepot{J. Le Potier, {\it Lectures on Vector Bundles}, Cambridge
University Press 1997}
\nref\lepotb{J.~Le~Potier, ``A Propos de la Construction de l'Espace de
Modules des Faisceaux Semi-Stables sur le Plan Projectif'', Bulletin de la
Soci\'et\'e Math\'ematique de France 0037-9484/1994/363}
\nref\geometric{W. Lerche, 
``Introduction to Seiberg-Witten Theory and its Stringy Origin,''
Nucl. Phys. Proc. Suppl. 55B (1997) 83-117; Fortsch.Phys. 45 (1997) 293-340;
hep-th/9611190.}
\nref\luta{M.A.~Luty and W.I.~Taylor, ``Varieties of vacua in 
classical supersymmetric gauge theories,'' Phys.\ Rev.\ {\bf D53}, 
3399 (1996) hep-th/9506098.}
\nref\mamost{M.~Mari\~no, R.~Minasian, G.~Moore and A.~Strominger,
``Nonlinear instantons from supersymmetric p-branes,'' 
JHEP {\bf 0001}, 005 (2000); hep-th/9911206.}
\nref\naknozak{M. Naka and M. Nozaki, 
``Boundary states in Gepner models,'' hep-th/0001037.}
\nref\okonek{C.~Okonek, M.~Schneider, H.~Spindler, {\it Vector Bundles
on Complex Projective Spaces}, Birkh\"auser Boston, 1980.}
\nref\rs{A.~Recknagel and V.~Schomerus, ``D-branes in Gepner
models," Nucl.Phys. B531, 185 (1998), hep-th/9712186.}
\nref\sardo{A.~V.~Sardo Infirri,
``Partial Resolutions of Orbifold Singularities via Moduli Spaces of 
HYM-type Bundles,'' alg-geom/9610004.}
\nref\schei{E.  Scheidegger, 
``D-branes on some one- and two-parameter Calabi-Yau hypersurfaces,'' 
hep-th/9912188.}
\nref\schofb{A. Schofield ``General representations of quivers,''
Proc. London Math. Soc. (3) 65 (1992) no.1, 46.}
\nref\schoa{A.~Schofield, ``Birational classification of moduli spaces
of representations of quivers'', math.AG/9911014}
\nref\schob{A.~Schofield, ``Birational classification of moduli spaces
of vector bundles over $\BP^2$'', math.AG/9912005}
\nref\seiwit{N. Seiberg and E. Witten, ``Monopole Condensation, And 
Confinement In N=2 Supersymmetric Yang-Mills Theory'', Nucl.Phys. B426 (1994) 
19-52; Erratum-ibid. B430 (1994) 485-486}
\nref\seidthomas{P. Seidel and R.P. Thomas ``Braid group actions on
derived categories of coherent sheaves,'' math.AG/0001043 and to appear.}
\nref\sharpedc{E. Sharpe, ``D-Branes, Derived Categories, and Grothendieck 
Groups'', Nucl.Phys. B561 (1999) 433}
\nref\thomas{R.P. Thomas, ``Derived categories for the working
mathematician,'' math.AG/0001045.}
\nref\tomasiello{A. Tomasiello, ``Projective resolutions of coherent sheaves 
and descent relations between branes'', hep-th/9908009.}
\nref\vafawit{C. Vafa and E. Witten,
``A Strong Coupling Test of S-Duality,''
Nucl. Phys. B431 (1994) 3-77, hep-th/9408074.}
\nref\linsig{E. Witten, ``Phases of N=2 Theories in Two Dimensions,''
Nucl. Phys. B403 (1993) 159, hep-th/9301042.}
\nref\smallinst{E. Witten, ``Small Instantons in String Theory,''
Nucl. Phys. B460 (1996) 541-559; hep-th/9511030.}
%
%
\newsec{Introduction}

D-branes in Calabi-Yau backgrounds of string theory have been studied
in a number of recent works, both to obtain the spectrum of BPS states
in these backgrounds and because their world-volume theories are
$d=4$ $\CN=1$ gauge theories which could appear in realistic models
and which naturally encode the geometry
and especially the moduli spaces of vector bundles on a CY.
Background and a list of references can
be found in \dreview; more recent works include 
\refs{\lerche,\naknozak,\schei,\brunscho,\lazaroiu,\gov}.

A primary question is to classify the BPS branes for a given CY.  Because
the bulk theory has $\CN=2$ supersymmetry, the spectrum of BPS branes will
depend on the moduli controlling the BPS central charges (the K\"ahler
moduli for B branes) and the classification must reflect this dependence.

In this work we begin the study of the spectrum of BPS branes
on the three complex dimensional ALE space $\CO_{\P^2}(-3)$.  The
stringy geometry of this space can be determined using local mirror
symmetry \refs{\mirror,\agm} and is quite similar to compact CY's such
as the quintic; the K\"ahler moduli space has a conifold point and a
point of enhanced discrete symmetry where the $\BP^2$ degenerates to a
$\BC^3/\BZ_3$ orbifold singularity.  Using these results, one can
relate D-brane charges at different points in moduli space \dg, as we
review in section 2.

This example has many advantages which will allow us to get a
fairly complete picture.  First, the spectrum of BPS branes in the
large volume limit follows from known mathematical results, namely
the classification of stable vector bundles on $\P^2$ \lepot.  
This classification proceeds in two
steps: one first constructs all holomorphic bundles on $\P^2$, and then
identifies a subset as stable.  We review the general construction of
holomorphic bundles on $\P^2$ due to Beilinson \beilinson, and related
topics in section 3.  

In section 4 we review results from the general analysis of D-branes
on orbifolds \dgm.  This analysis provides quiver gauge theories
describing all BPS branes at and near the orbifold point as
supersymmetric vacua.  The problem of finding such vacua also proceeds
in two steps: first, find F-flat configurations, then minimize the D
term contributions to the potential.
As it turns out, there is a very relevant branch of mathematics, the theory
of quiver representations (we provide an introduction in the
appendix), in which the problem again becomes to find all
``holomorphic objects'' and then find the stable subset.  
This will allow us to give a good picture of the spectrum of branes at
and near the orbifold point.

It will turn out that the holomorphic objects at the two points in
moduli space are essentially the same, in a sense we describe in
detail.  The point is that the orbifold quiver theory turns out to
reproduce Beilinson's general construction of bundles on $\BP^2$.
This allows us not only to identify D-brane charges with Chern classes
of bundles, but to identify the entire moduli spaces of these
configurations and even explicitly construct the bundle corresponding
to any point in the quiver theory moduli space.

In section 5 we compare the BPS spectrum in the large volume limit
with that in the neighborhood of the orbifold point.  In both cases,
the BPS spectrum consists of stable holomorphic objects, but with
a different definition of stability in the two regimes:
$\mu$-stability in the large volume limit, and $\theta$-stability
\king\ near the orbifold point.  We describe the spectrum near the
orbifold point in some detail; it is very dependent on the particular
line coming in, and on some lines (such as the one to the conifold
point) is significantly smaller than the large volume spectrum
but still infinite.

Both of these definitions of stability are special cases of a proposed
condition determining the BPS branes at general points in moduli space
\pistability.  This starts from the ``decoupling conjecture'' of
\bdlr.  For B branes, this states that the holomorphic structure
(field content and F-flatness conditions) of such a world-volume
theory depends only on the complex structure of the CY, while the
D-flatness conditions depend only on the K\"ahler moduli.  Further
discussion and arguments for these claims can be found in \dtoappear.

Given decoupling, we can assert that the problem of finding BPS branes
at any point in moduli space can be studied by the same two-step
procedure: identify a category of holomorphic objects, which at least
locally does not depend on K\"ahler moduli, and then for a specific
point in K\"ahler moduli space identify the stable subset.  (See
\dtoappear\ for an explanation of the relevance of the term
``category'' here.)  In \pistability\ a simple criterion (called
$\Pi$-stability) was proposed, which reduces to the known physical
stability conditions in all examples studied so far.

In section 6 we apply $\Pi$-stability to develop a picture of how the
BPS spectrum evolves from the orbifold point to the large volume
limit.  It will turn out that there are
an infinite number of lines of marginal stability; we will show how 
a specific line or brane decay could be studied.
We will also exhibit lines of marginal stability arbitrarily close
to the large volume limit.  We attempt to incorporate all this
information in a general picture of the spectrum.

Beilinson's construction of holomorphic bundles applies to 
projective space of any dimension, and the results of \bdlr\ for
D-branes on the quintic bear a striking relation to the construction
for $\P^4$, which we explain in section 7.

Section 8 contains conclusions.

Some general points of notation and convention.
We are studying the problem of brane configurations in type \II\ string
theory in the classical limit, zero string coupling but non-zero string 
length, described by CFT with sphere and disk world-sheets.
We always refer
to branes by the dimension of the cycle they wrap in the CY, but
usually describe their world-volume theories in terms of
the $d=4$, $\CN=1$ supersymmetric gauge theory which would be obtained
by letting them extend in the $3+1$ Minkowski dimensions.

\newsec{The stringy moduli space of $\CO_{\P^2}(-3)$}

We are interested in a noncompact CY $\CM$ which can be defined in two
ways.  One way is as a line bundle over the complex projective plane
$\P^2$, whose first Chern class $\chcl_1=-3$ is chosen to produce
$\chcl_1(\CM)=0$.  The homology is that of $\P^2$ with $b_0=b_2=b_4=1$; let
$\omega$ be the generator of $H^2(\BZ)$. The Ricci flat metric on this
space was written down explicitly by Calabi \calabi\ and should be a good
description for large K\"ahler class (in string units).  In this metric
it is clear that compact minimal volume surfaces embed entirely into
the $\P^2$.  

The same space can be defined as the blowup of the orbifold
singularity $\BC^3/\BZ_3$ and indeed Calabi's metric is asymptotically
locally Euclidean (ALE) to this flat geometry.  On the other hand, for
small K\"ahler class the curvature of the $\P^2$ is large, so in this
regime the precise form of the metric and other geometric predictions
cannot be trusted a priori.

The stringy moduli space is a Riemann sphere with three punctures.
One of these is the large volume limit for which a good coordinate
is $z=\exp 2\pi i (B+iJ)$ where $B+iJ$ is the complexified K\"ahler
form. In this coordinate system the large volume limit is at $z=0$.
The observables of most interest for us are the masses of
BPS $2k$-branes which in this limit will be 
\eqn\masslimit{
m_{2k} = |\Pi_{2k}| \rightarrow |{1\over k!}(B+iJ)^k|.
}
These receive world-sheet instanton corrections which can be expressed
as a Taylor series in $z$ with radius of convergence $1$ determined by
a singularity with logarithmic monodromy at $z=1$, the conifold point.
Further analytic continuation reaches
a point ($z=\infty$) with $\BZ_3$ monodromy, the orbifold point.

The periods $\Pi_{2k}$ can be computed using local mirror symmetry
\refs{\agm,\mirror}\ and satisfy the following differential equation
(where $\theta_z=z{d\over dz}$):
\eqn\perioddgl{
\left[\theta_z^3-z(\theta_z+{1\over 3})
(\theta_z+{2\over 3})\theta_z\right]\Pi=0.
}
There are three linearly independent solutions (including $\Pi_0=1$).

Let us define a basis $\Pi_{2k}$ in which the mass of a brane
with charges $Q_{2k}$ is $m=|Q\cdot\Pi|$.
We define these charges in terms of the topological class of
a D$4$-brane configuration: this is determined 
by the number of branes $N$ wrapping $\P^2$ and the Chern
character $\tr e^{F}$
of the vector bundle $E$ it carries (which is a source of RR
charge in the usual way).  They are then
\eqn\chargedef{
Q_4=\rk(E)=N; \qquad Q_2 = \int_\Sigma \chcl_1(E); \qquad
Q_0 = \int \chch_2(E).
}
(note that charges are not integral in this basis).

One property of this basis is that
the large volume monodromy $B\rightarrow B+1$ is
\eqn\lvmona{
M_\infty^{(LV)}=\left(\matrix{1&0&0\cr 1&1& 0\cr \half& 1&1}\right)
}
acting (on the right) on the charge vector $(Q_0\ Q_2\ Q_4)$.
(it has the same effect as the shift 
$\int_\Sigma F\rightarrow \int_\Sigma F+ 1$).
This operation corresponds to tensoring the vector
bundle $V$ with a line bundle, an operation 
which (at large volume) preserves the attributes of the brane
(stability and the dimension of the moduli space) which enter our discussion.

Asking that this monodromy acts in the corresponding way on the periods
determines the basis up to a monodromy of
the same form and up to an overall constant shift of $\Pi_4$
(in terms of the charges, there is an ambiguity $Q_0\rightarrow Q_0-c Q_4$).
A basis for which this is true is defined in \dg: it is\foot{Note that what
we call $t$ is $t_b$ in \dg, also note that $t$ can sometimes be viewed as
a two form.} $1$, $t$ and $t_d$ defined by the large volume
asymptotics $t \sim B+iJ$ and $t_d \sim \half t^2 + {1\over 8}$.

This determines $\Pi_2=t$ and $\Pi_4=t_d+c$ for some constant $c$.
One can determine $c=0$ by \refs{\dg,\diro} matching the asymptotic formula
\eqn\zchargea{
Z(E)=-\int e^{-t} \chch(E) \sqrt{\aroof(T)\over \aroof(N)},
}
where $t$ is the complexified K\"ahler form and $T$ and $N$ are the 
tangent and normal bundles of the blown up $\BP^2$. 
One can also check the mass formula -- curvature terms in the D$4$ world-volume
action \bachas\ produce a mass shift which is $1/8$ the D$0$ mass.
These two determinations are surely related by supersymmetry.

We turn to the other singular points.
In this charge basis, the monodromy around the conifold point is
\eqn\conimona{
M_c^{(LV)}=\left(\matrix{1&0&0\cr 0&1& -3\cr 0&0&1}\right),
}
with $\Pi_4=t_d=0$ and $t = - 3 t_d \log(1-z) + O(1)$.
This is the usual monodromy associated with a massless BPS state,
here the pure (trivial vector bundle) D$4$-brane.

The orbifold point has $\BZ_3$ monodromy; in this basis
\eqn\orbmona{
M_o^{(LV)}=M_c^{(LV)} M_\infty^{(LV)} =
\left(\matrix{1&0&0\cr -\half&-2& -3\cr \half& 1&1}\right).
}
Here $t_d=1/3$ and $t=-\half$.

Considerations at this point are simpler in terms of a basis 
consisting of the D$4$-brane and its $\BZ_3$ images.  These will turn
out to be the elementary ``fractional branes,'' as was shown in
\dg\ and as we discuss in section 4.
Their periods have the asymptotic form\foot{
We changed the numbering of the branes from \dg, for reasons which
will become clear.}
\eqn\zchargeb{
\Pi_1=\half t^2+t+{5\over 8}+\cdots,\qquad
\Pi_2=-t^2-t+{1\over 4}+\cdots,\qquad
\Pi_3=\half t^2+{1\over 8}+\cdots.
}
Comparing \zchargea\ and \zchargeb\ we can express the large volume charges
$(Q_4\ Q_2\ Q_0)$ in terms of the orbifold charges $(n_1\ n_2\ n_3)$
\eqn\lvorbbasis{
Q_0 = -{n_1+n_2\over 2};\qquad Q_2=n_1-n_2;\qquad Q_4=-n_1+2n_2-n_3;
}

The monodromy matrices in the fractional brane basis are
\eqn\orbmonb{
M_o^{(O)} = \left(\matrix{0& 1& 0\cr 0&0&1\cr 1&0&0} \right),
}
\eqn\conimonb{
M_c^{(O)} = \left(\matrix{1& 0& -3\cr 0&1&3\cr 0& 0& 1}\right)
}
and
\eqn\lvmonb{
M_\infty^{(O)} = \left(\matrix{0& 1& 0\cr -3& 3& 1\cr 1& 0& 0}\right).
}

Let $L$ be the negative $\xi={1\over z}$ axis, a line connecting the orbifold
point $\xi=0$ to the large volume limit $z=0$ which intersects no 
singularities (it is opposite to the conifold point).

\ifig\periodplot{The evolution of the periods along the negative $\xi$
axis}{\epsfxsize2.0in\epsfbox{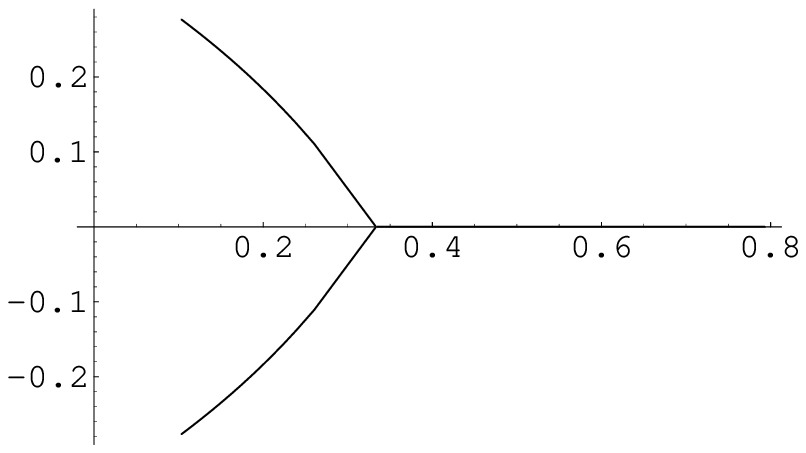}}

The evolution of the central charges of the three fractional branes 
along $L$ away from the orbifold point is plotted 
in \periodplot. All three periods are $1\over 3$
at the orbifold point. The periods $\Pi_1$ and $\Pi_3$ 
are complex conjugate, while the period 
$\Pi_2$ increases along the positive real axis. 

Note in particular that the relative periods encircle zero, which is
the basic criterion for significant changes in the BPS spectrum (and
other physics) between the two limits.

Note also that any charge vector for which $\sum n_i = 0$ potentially
describes a BPS brane which would become massless at the orbifold point.
Many of these branes are known to exist in the large volume limit, the
simplest example being $(1\ 0\ -1)$ which is the two-brane with $c_1=0$
(no magnetic flux).  This might be a little surprising as there is
a sense in which $B=\half$ at the orbifold point \agm\ (this was our
earlier $t=-1/2$).  However, 
the other world-volume couplings conspire to cancel
this contribution to the mass.

Assuming the theory is consistent, because the
orbifold CFT is non-singular, all of these branes will have to decay
before reaching the orbifold point.  Thus we can already see some
significant differences between the large volume and stringy spectrum.

\newsec{Branes in the large volume limit}

In this section we describe the spectrum of BPS branes in the large
volume limit.  We first note that in this paper we will only discuss
the classical (zero string coupling)
spectrum; in other words branes which can be defined
as boundary states in conformal field theory.
We summarize some standard notations from algebraic geometry in appendix B.

The branes in the large volume limit fall into the following three
classes:

\item{$\bullet$}
The D$0$-brane, with $Q_0=1$, $Q_2=Q_4=0$.

\item{$\bullet$}
D2-branes ($Q_4=0$) are described by holomorphic and antiholomorphic
curves in $\P^2$ (characterized by degree) carrying a vector bundle
(characterized by $\chcl_1$).  The homogenous equations $x^n+y^n+z^n=0$ in
$\BP^2$ represent all positive integral degrees and thus all integral
$Q_2\ne 0$ appear ($Q_2<0$ appear as antiholomorphic curves).  They
are curves of genus $1+n(n-3)/2$ and can carry line bundles of
arbitrary integral $\chcl_1$; thus all $Q_0$ can appear.
\eqn\dtwocharges{
n=Q_2,\qquad \chcl_1=Q_0.
}

\item{$\bullet$}
D4-branes are described by $\mu$-stable holomorphic vector bundles
on $\P^2$, characterized topologically by the rank $\rk$, $\chcl_1$ and
$\chcl_2$.  These are related to $Q_{2k}$ as $Q_4=\rk$, $Q_2=\chcl_1$ and
$Q_0=\chch_2$.

Although every triple $(\rk,\chcl_1,\chcl_2)$ with $\rk>0$ appears as
a topological bundle, not all of these can support stable bundles.
The importance of this condition for us comes from Donaldson's theorem
on the existence of solutions to the anti-self-dual Yang-Mills
equation 
\donaldson.
An unstable bundle will not admit a solution,  and thus any such
four-brane will not be BPS.  A semistable bundle will only
admit reducible self-dual connections, which admit more than one
independent covariantly constant scalar $D_i X=0$.  These include
moduli for motion in $\BR^{3,1}$ and thus these
configurations can be broken up into several branes.

In the following, instead of discussing holomorphic vector bundles, we
will be a bit more general and use coherent sheaves, mostly because
these behave much better in many of the mathematical constructions.
Physically, this more or less corresponds to including subbranes
supported on submanifolds of the original brane and is useful for
describing point-like instantons and quantum bound states
\hm.  We also note that, although it is not a priori obvious, it turns
out that for $\P^2$, a stable coherent sheaf at generic points in its
moduli space is in fact a bundle \lepot.

\subsec{Stable bundles on $\P^2$}

We proceed to quote the classification of stable bundles on $\P^2$.
The two invariants which determine stability are the
{\it slope}
\eqn\slopea{
\mu = {\chcl_1\over \rk} = {Q_2\over Q_4}
}
and {\it discriminant}
\eqn\discra{
\Delta = {1\over 2\rk^2} (2\rk \chcl_2 -(\rk-1)\chcl_1^2) = 
{1\over 2 Q_4^2}(Q_2^2-2Q_0Q_4).
}
(We will sometimes refer to D0 and D2 branes as having 
infinite slope and discriminant.)

A sheaf $E$ is $\mu$-stable (respectively $\mu$-semistable) if 
it is torsion-free\foot{
This more or less means that no D$p-2$-subbranes appear.}
and for every coherent sub-sheaf $E'$ of rank $0<\rk'<\rk$,
we have $\mu(E') < \mu(E)$ (respectively $\le$).
The geometric origin of this condition is explained
in \refs{\donaldson,\friedman,\lepot}.
The basic idea of Donaldson's theorem is to find a solution of the
Yang-Mills equations as a minimum of the action on an orbit of the 
complexified gauge group.  Such a minimum will exist if the orbit is
closed; conversely it can be shown that if the orbit is not closed,
the minimum lies off of the orbit and no solution exists.
Stability is the necessary and 
sufficient condition for the orbit to be closed.  

It is interesting to note that the condition
$\mu(E')<\mu(E)$ on $\P^2$
can be expressed in terms of an intersection form
on $\BC^3/\BZ_3$: it is
\eqn\chargestab{
\sgn (Q_4 Q_4') (Q_2' Q_4-Q_2 Q_4') < 0.
}

The discriminant appears in Bogomolov's inequality
\eqn\bogol{
{\rm semistability} \Rightarrow \Delta \ge 0
}
as well as in the formula for the expected complex dimension of moduli space,
\eqn\lvdim{
d = H^1(E^*\otimes E)=1 + \rk^2(2\Delta - 1).
}
Bogomolov's inequality can be proven by assuming
an anti-self-dual Yang-Mills connection $F$ exists and doing a
straightforward computation using the inequality (for hermitian $M=iF$) 
$(\tr M)^2 \le N \tr M^2$ \friedman.
It is a necessary but not sufficient condition for this connection
to exist.  The expected dimension can be computed using the index theorem
and it is a non-trivial result of the theory that, when the connection
exists, the moduli space is (locally)
a manifold of the expected dimension.  Thus $d\ge 0$ gives a stronger
necessary condition.

In fact there are two cases:

\item{$\bullet$}
$d=0$ (then $\Delta<\half$).  
This is an ``exceptional bundle.''  It can be shown that, for a given
slope $\alpha$, there exists at most one exceptional bundle, and its 
rank is the least positive integer such that $\rk\alpha\in\BZ$. 
Then \lvdim\ determines $\chcl_2$ and $\Delta_\alpha$.
Let $\CE$ be the set of slopes of exceptional bundles; we will
describe it in section 5.

\item{$\bullet$}
$d>0$ (then $\Delta\ge\half$) are ``regular bundles.''
The additional necessary condition for this bundle to exist is
\hfill\break
(*)
for all $\alpha\in\CE$ such that $\rk_\alpha<\rk$ and $|\mu-\alpha|<3$,
we have
\eqn\addcond{
\Delta+\Delta_\alpha \ge P(-|\mu-\alpha|),
}
where the Hilbert polynomial is given by $P(\nu)=1+\half(\nu^2+3\nu)$.

This condition follows directly from the $\mu$-stability condition and
the Riemann-Roch formula:
\eqn\defchi{\eqalign{
\chi(E_1,E_2) &\equiv \sum_i (-1)^i \dim_\BC \Ext^i(E_1,E_2) \cr
&= \rk_1 \rk_2 (P(\mu_2-\mu_1) - \Delta_1 - \Delta_2).
}}
Suppose $\alpha>\mu$; then a stable bundle $E$ of slope $\mu$ must satisfy
$H^0(E_\alpha,E)=0$ (or else the image of such a morphism would be a 
destabilizing subbundle, see section 6). Similarly if $\alpha-\mu<3$, then 
$\Ext^2(E_\alpha,E)=0$, or else Serre
duality tells us that $H^0(E_{-\alpha},E^*\otimes \Omega^{-1})\ne 0)$
destabilizes a related stable bundle.  This tells us that 
$\chi(E_\alpha,E)\le 0$ which translates into \addcond.

We have now given a set of conditions which are clearly
necessary for stability.  They are also sufficient, but the proof is rather
complicated and we refer to \lepot. They can also be made much more concrete, 
but we postpone this discussion to section 5.

\subsec{Translation into orbifold basis}

It will be useful to have these formulas translated into the orbifold basis.
The slope, discriminant and the expected dimension of the moduli space
then take the form
\eqn\lvorbbasisb{\eqalign{
\mu &={{n_2-n_1}\over{n_1-2n_2+n_3}} \cr
\Delta &= {-n_1n_2-n_1n_3-n_2n_3 + 3n_2^2 \over 2(n_1-2n_2+n_3)^2} \cr
d &= 1 - \half n^t\cdot C^{LV}\cdot n
}}
with the ``Cartan'' matrix
\eqn\cartanlv{
C^{LV} = \left(\matrix{2& -3& 3\cr -3& 2& -3\cr 3& -3& 2}\right).
}
Exceptional bundles satisfy $n^t\cdot C^{LV}\cdot n = 2$, while stable 
nonexceptional bundles $E$ must satisfy 
$n^t\cdot C^{LV}\cdot n \le 0$ and the second condition above:
for every $\alpha\in\CE$ with $|\alpha-\mu|<3$,
$\chi(E_\alpha,E) \le 0$ where
\eqn\chiorbbasis{
\chi(E,E') = n\cdot 
\left(\matrix{1&-3&3\cr 0& 1& -3\cr 0& 0& 1}\right)
\cdot n'.
}

\subsec{Beilinson monads}

Sheaves on $\P^2$ and their moduli spaces
can be described explicitly by use of
``Beilinson monads'' \refs{\beilinson,\lepot,\okonek,\huleka,\hulekb}.
This is much like the linear sigma model construction
of vector bundles popular in the physics literature \linsig.

We consider a quiver (see appendix) with three vertices $V_1$, $V_2$ and
$V_3$ and six arrows: $X^i$ for $1\le i\le 3$ run from $V_1$ to $V_2$,
and $Y^i$ for $1\le i\le 3$ run from $V_2$ to $V_3$
We furthermore impose the relations
\eqn\quiverrel{
Y^i X^j = Y^j X^i.
}
\ifig\beiquiver{The Beilinson quiver}{\epsfxsize1.5in\epsfbox{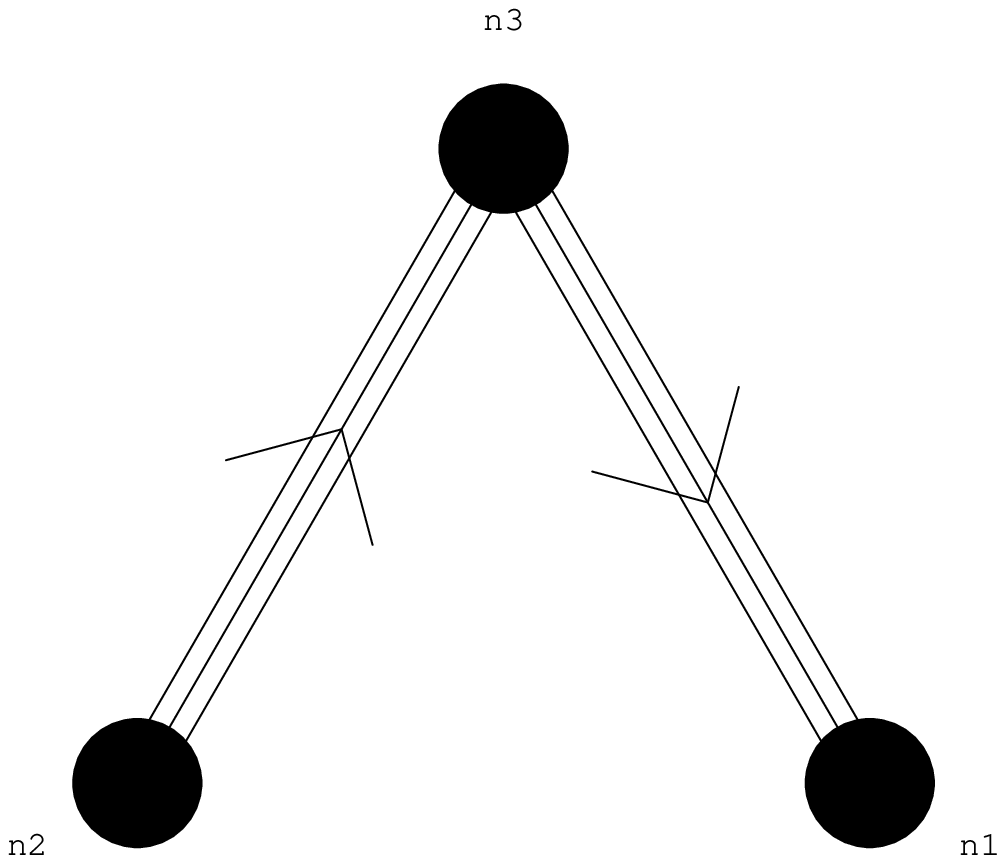}}

As is easy to check, the expected dimension of the moduli space
of such a quiver is
\eqn\lvdimformula{d = 1 - \half n^t\cdot C^{LV}\cdot n}
where $n_i=\dim V_i$, exactly the dimension of the moduli space of bundles
on $\P^2$ expressed in the basis \lvorbbasis.
This is not a coincidence -- the moduli spaces of these quivers
and the moduli space of bundles with these charges are the same.

In fact, there is a detailed correspondence between representations of
this quiver and coherent sheaves on $\P^2$, which we now explain.
A representation 
$(n_1\ n_2\ n_3)$ of the quiver can be rewritten as a complex, 
\eqn\quivercomplex{
0\rightarrow \BC^{n_1} \otimes \Lambda^2V^*\mapr^X
\BC^{n_2} \otimes V^* \mapr^Y \BC^{n_3} \rightarrow 0,
}
where $V=\BC^3$ with basis $(e_i)$. The maps are the
natural contractions with $X^i e_i$ and $Y^i e_i$.
The relations of the quiver are then equivalent to $Y\cdot X=0$.

We take $\BP^2=\BP(V)$, so the usual projective coordinates $z^i$ on
$\P^2$ are elements of $V^*$ (linear functions on $V$).  They are also
sections of $\CO(1)$, so we can write the exact sequence (dual to the
Euler sequence \gh)
\eqn\eulersequ{
0\rightarrow \Omega(1) \rightarrow \CO\otimes V^* \mapr^{*} \CO(1) 
 \rightarrow 0
}
where $\mapr^{*}$ is the evaluation map (multiplication of the two
factors).  Translating into components, this just says that $\Omega$,
which is the cotangent bundle on $\P^2$, has local sections which can
be written as $\psi_i(z) dz^i$ satisfying $z^i \psi_i(z) = 0$.

Thus, we can tensor \quivercomplex\ with $\CO$ and restrict
$\CO\otimes \Lambda^iV^*$ to $\Omega^i(i)$ (and use 
$\Lambda^2\Omega(2) \cong \CO(-1)$), to get a new complex
\eqn\monad{
0\rightarrow \BC^{n_1} \otimes \CO(-1)\mapr^{\hat X}
\BC^{n_2} \otimes \Omega^1(1) \mapr^{\hat Y} \BC^{n_3} \otimes \CO
 \rightarrow 0,
}
with maps $\hat X$ taking $\psi_{[ij]}$ to $X^j\psi_{[ij]}$ and 
$\hat Y$ taking $\psi_i$ to $Y^i \psi_i$.

This complex is the Beilinson monad.  If we start with
a Schur representation of the Beilinson quiver, this sequence is exact at 
the  first and third node, and the cohomology at the second node is
a simple coherent sheaf $E$ with Chern character
\eqn\cohomologychern{
\chch(E) =n_2\chch(\Omega^1(1))-n_1\chch(\Omega^2(2))-n_3\chch(\CO).
}
One can check that
\eqn\basicchern{
\chch(\CO(-1))= 1-\omega+\half\omega^2,\qquad
\chch(\Omega(1))=2-\omega-\half\omega^2,\qquad
\chch(\CO)=1
}
so that
\eqn\cohomologycherntwo{
\chch(E) = Q_4+Q_2\omega+Q_0\omega^2
} 
where we take the relation between $n_i$ and $Q_{2p}$ to be exactly
\lvorbbasis. This match between the result from mirror symmetry and 
the Beilinson construction is quite remarkable and gives hope for similar 
constructions.

\medskip

We now discuss how to obtain a quiver from a sheaf $E$.
The simplest case is to suppose its slope
satisfies $-1<\mu\le 0$ (this is sometimes called a ``normalized'' sheaf,
and can be arranged by tensoring with a line bundle).
Under this condition, one can show that 
$h^0(E(n))=h^2(E(n))=0$ for $-2\le n\le 0$, and applying
the Riemann-Roch theorem one finds
\eqn\homologies{\eqalign{
n_3 &= h^1(E)=-\chi(E)=\rk(E)+{3\over 2}\chcl_1(E)+\chch_2(E),\cr
n_2 &= h^1(E(-1))=\chcl_1(E)+\rk(E)-\chi(E),\cr
n_1 &= h^1(E(-2))=2\chcl_1(E)+\rk(E)-\chi(E)
}}
which is exactly the inverse of the relation \cohomologycherntwo\ above.

The quiver is then obtained by taking $H^1(E(-2))$, $H^1(E(-1))$ and $H^1(E)$
as representation spaces for the vertices.  The maps $X^i$ and $Y^i$ are the
restrictions to cohomology of the action of multiplication by $z^i$ on a
section of $E(n)$, giving a section of $E(n+1)$.
These maps obviously satisfy the relations $Y^i X^j=Y^j X^i$. 

There is then a theorem of
Beilinson that states that the cohomology of the corresponding monad \monad\
is exactly the sheaf $E$. The proof, although short,
uses a good deal of homological algebra, which we will not get into
here (see \lepot, pp. 182--183).

This goes a long way towards establishing a one-to-one relation
between quiver representations and sheaves, but the relation we just
described is not completely general.  It can hold only when the
charges satisfy the inequality
\eqn\rankinequ{
\rk=2n_2 - n_1 - n_3 > 0
} 
as otherwise $\ker v/\im u$ is degenerate.  This is guaranteed if
the sheaf was normalized, so this suffices to construct all moduli
spaces (since these are invariant under tensoring with line bundles),
but does not give a general relation.

Beilinson's theorem in fact applies to all sheaves and states that the
bundle $E$ can be recovered as the cohomology of a spectral sequence
with initial term
\eqn\initerm{
E^1_{p,q} = H^q(\P^2, E(p)) \otimes \Omega^{-p}(-p).
}
If $h^q\ne 0$ for a single $q$, this will reduce to the cohomology of
a complex, as we described.  This provides a partial generalization of
the relation, to the cases where only $h^0\ne 0$ (true after for
tensoring with $\CO(n)$ for some $n>>0$), and only $h^2\ne 0$ (true
for some $n<<0$), which will also be used below.

\newsec{Branes at the orbifold point}

\subsec{The $\BC^3/\BZ_3$ quiver gauge theory}

The construction of \refs{\dm,\dgm} 
produces the world-volume theories describing 
configurations of any set of branes which can become BPS at or near
the orbifold point. These are quiver gauge
theories $(n_1\ n_2\ n_3)$ labelled by three integers $n_i$:

\ifig\quiver{The $\BZ_3$ quiver}{\epsfxsize1.5in\epsfbox{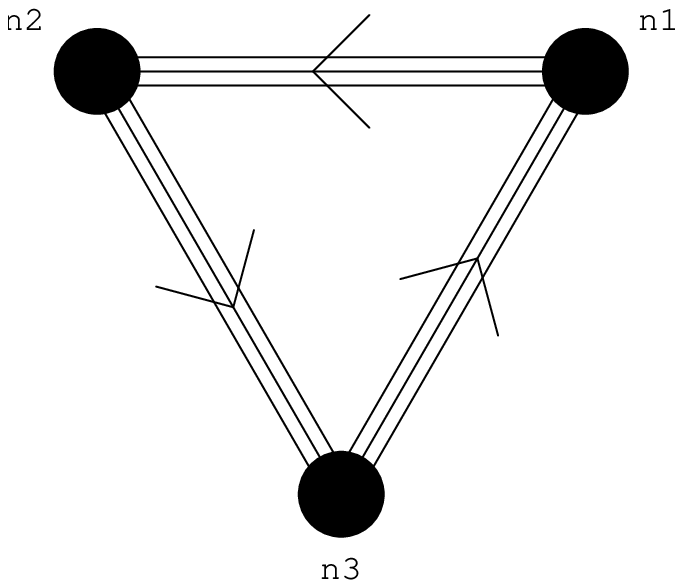}}

The theory has gauge group
$U(n_1) \times U(n_2) \times U(n_3)$ and matter content
$3(n_1,\bar n_2)+3(n_2,\bar n_3)+3(n_3,\bar n_1)$.
Let the chiral superfields be notated $Z^i_{a,a+1}$ with $a\in\BZ_3$;
there is a superpotential
$W=\epsilon_{ijk} \tr Z^i_{12} Z^j_{23} Z^k_{31}$.  
The gauge group contains three $U(1)$ factors, one of which acts trivially,
allowing for two independent
Fayet-Iliopoulos terms; these are controlled by the
closed string moduli (which also have real dimension $2$).
There is a $\BZ_3$ symmetry acting on the set of theories by cyclically 
permuting the $n_i$'s and the associated FI terms.

The basic result \refs{\dgm,\sardo}\ is that the theory $(1\ 1\ 1)$ 
describes a D$0$-brane moving on
$\CM$.  The Higgs branch of its moduli space 
is generically a non Ricci-flat ALE space
with $c_1=0$ and K\"ahler class depending on the FI terms.
The $\BZ_3$ symmetry acts on the FI terms in a way consistent with identifying
it with the $\BZ_3$ monodromy of the orbifold point.

The other theories describe different sectors in string theory.
One can compute the twisted Ramond-Ramond charge and find it is non-zero
for $n_i\ne n_j$, so these are branes wrapped about the exceptional
cycle or ``fractional branes.''
The elementary theories are $(1\ 0\ 0)$ and its images under $\BZ_3$.
Since these all preserve
the same supersymmetry and sum to the D$0$,
all of their central
charges are equal to $1/3$; combining this information
with the CFT intersection form fixes the
identification of the indices $n_i$ with the orbifold charge
basis defined in section 2 \dg.
In particular, $(1\ 0\ 0)$ is the ``pure''
D$4$-brane (the state which becomes massless at the conifold point).

The definition of a bound state is a gauge theory $(n_1\ n_2\ n_3)$
with a Coulomb branch corresponding to that of a single object in
the Minkowski dimensions.  This will be true if the unbroken gauge symmetry
is precisely $U(1)$.

The question of which combinations of $n_i$ correspond to bound states has
a priori different answers in the classical and quantum theory.
We will discuss the classical theory, in which the theory
$(n_1\ n_2\ n_3)$ describes a bound state if it admits a gauge equivalence
class of vacua (satisfying the superpotential and D-flatness conditions)
which break the gauge symmetry to $U(1)$.  
As we discuss in the appendix, such configurations
correspond to ``semistable Schur representations'' of the quiver
with relations.

We now claim that, with the single exception of the theory
$(1\ 1\ 1)$ (the D$0$), such representations of the
$\BZ_3$ quiver always come from representations of the Beilinson quiver.
In other words, the $Z_{a,a+1}^i=0$ for one $a\in\BZ_3$ and all $i$.
The precise relation depends on which region of moduli space we
consider; let us consider the region with $\zeta_3 > 0$ and $\zeta_1< 0$.

Looking at states with all $n_i\ne 0$ we need to solve the additional
F flatness constraints: $Z^i Z^j = Z^j Z^i$ for all three pairings.
Naively the expected dimension for this quiver is always negative;
this is too naive because the F flatness constraints will always be redundant.
For example, in the case $(1\ 1\ 1)$ four conditions are redundant leading
to the dimension $3$ for the D$0$ moduli space.

An easy way to eliminate most of these constraints is to set $Z_{a,a+1}=0$
and reduce to the Beilinson quiver, which we know has Schur representations.
In the region of moduli space under discussion, this produces a Beilinson
quiver with $n_2$ at the middle vertex and thus the discussion of the previous
section tells us for all charge vectors whether or not
Schur representations exist.

We have not proven that this exhausts the Schur representations.
However, we have two reasons for believing this is so.
Physically, taking all $Z_n\ne 0$ allows the brane to leave the
exceptional divisor (the $\BP^2$), and the only BPS brane we expect to
do this is the D$0$.

A purely mathematical argument for this point might go as follows.
First, if we had set $Z^2_{n,n+1}=Z^3_{n,n+1}=0$,
the single matrix $Z^1_{n,n+1}$ could only break the gauge
symmetry to at least $U(1)^n$ where $n=\min(n_1,n_2,n_3)$.

Then, generic solutions with $Z^2\ne 0$ and $Z^3\ne 0$ do not break any
more gauge symmetry.  This can be seen by assembling the
$Z^1$, $Z^2$ and $Z^3$ into the original matrices of the underlying
$\CN=4$ theory (satisfying $\gamma^{-1} Z^i \gamma = \omega Z^i$),
for which the F flatness conditions are $[Z^i,Z^j]=0$.
The solutions of these are then
\eqn\fflatsln{
Z^2 = f^2(Z^1); \ Z^3 = f^3(Z^1) 
}
which break the same gauge symmetry as $Z^1$.

This leaves only the possibility of non-Beilinson representations with
some $n_i=1$.  We have looked at the examples $(1\ 1\ n)$ and $(1\ 2\ 2)$
in detail and found that these solutions do not break to $U(1)$;
taking larger $n_2$ and $n_3$ drives the expected dimension negative and
thus seem unlikely to lead to Schur representations.

\medskip 

The conclusion of this section is that holomorphic objects near the
orbifold point, with the exception of the D$0$, come from
representations of the Beilinson quiver.  As we saw in the previous
section, each such representation is in precise correspondence
to one particular large volume sheaf (it would be interesting to
see this more physically, perhaps by using a D$0$-probe).
Thus in this
theory we can make a very detailed identification between many of the
holomorphic objects at very distant points in K\"ahler moduli space.
This seems to us to be striking evidence for a strong form of the
``decoupling conjecture'' of \bdlr, that the holomorphic objects are
independent of K\"ahler moduli.  On the other hand, the two sets of
holomorphic objects are not literally identical, but we defer
discussion of this point to section 6.

\newsec{BPS branes in the large volume limit and near the orbifold point}

We are now in a position to compare the BPS spectra near these two
points.  Based on the behavior of the periods and an analogy to the
case of pure $\CN=2$ gauge theory \seiwit\ it was suggested in
\dreview\ (for the quintic, but the situation here is very similar)
that the spectrum in the stringy regime could be substantially smaller
than the large volume spectrum.  In fact  the spectrum near the
orbifold point is strongly dependent on the direction we approach it from,
and we will see in what sense this suggestion is true.

We start by completing the discussion of stable bundles on $\BP^2$  \lepot.
The basic criterion to get any representation at all is that the
expected dimension of the moduli space (\lvdim\ or \lvorbbasisb) $d \ge 0$.
The regular solutions have $d>0$ or equivalently $\Delta>\half$, and
this part of the spectrum is approximately described by a single
quadratic inequality,
\eqn\regularquad{
Q_2^2 - 2Q_0Q_4 - Q_4^2 > 0.
}
This is however not the precise description for two reasons, which
will lead to finer structure near the boundary of \regularquad.
First, $d=0$ for an infinite discrete set of ``exceptional bundles''
with $0<\Delta<\half$.  Then, we need to check $\mu$-stability for all
of these bundles, using \addcond.

We proceed to quote this precise result in the next section.
First, we quote the set of exceptional bundles $\CE$.  Then,
one can show \lepot\ that 
for bundles of given slope $\mu$, a single exceptional 
bundle with slope $\alpha$ dominates \addcond, reducing it to a 
single inequality
\eqn\stabcond{
\Delta \ge \delta_\alpha(\mu).
}

\subsec{Exceptional bundles and stable bundles on $\P^2$}

We now describe the set $\CE$.  First, by considering tensor products
with line bundles and dualization, one sees that 
$\CE \cong -\CE \cong \CE + 1$. Starting with $\BZ\subset\CE$, one 
can produce the remaining elements by the following inductive procedure.
Let $\rk_\alpha$, $\chi_\alpha$ and $\Delta_\alpha$ be the invariants 
associated to an exceptional bundle of slope $\alpha$ as above.

It is convenient to parameterize points in the set $\CE$ by fractions of
the form $p/2^q$.  So, let $\epsilon(p/2^q)$ be an increasing function 
from the set of such fractions
to the set $\CE$, with $\epsilon(n)=n$ for $n\in\BZ$.
Then
\eqn\epsdef{
\epsilon({2p+1\over 2^{q+1}}) = 
	\epsilon({p\over 2^q}) * \epsilon({p+1\over 
2^q}) 
}
where 
\eqn\starprod{
\alpha * \beta = {\alpha+\beta\over 2} + 
    {\Delta_\beta-\Delta_\alpha\over 3+\alpha-\beta}.
}
The first few examples are 
\eqn\epstable{
\matrix{{p\over 2^q} & 0 & {1\over 8} & 
{1\over 4} & {3\over 8} & {1\over 2} \cr\cr
\epsilon({p\over 2^q}) & 0 & {5\over 13} & 
{2\over 5} & {12\over 29} & {1\over 2}}
}
and they rapidly accumulate to $\Delta\rightarrow 1/2$.

Using this description, it can be shown that
the condition \addcond\ follows from a simpler condition on $\Delta$
and $\mu$.
We define the function $\delta(\mu)$ as
\eqn\deltadef{
\delta(\mu) = \sup_{\alpha\in\CE; |\mu-\alpha|<3} 
P(-|\alpha-\mu|)-\Delta_\alpha.
}
This function satisfies $\half < \delta(\mu)<1$.
Non-exceptional sheaves then satisfy \stabcond.

It is useful to give a geometric construction of $\CE$ and
$\delta(\mu)$. 
Given two slopes $\alpha$ and $\beta$, one can plot the functions
$\delta_\alpha(\mu)=P(-|\mu-\alpha|)-\Delta_\alpha$ and $\delta_\beta(\mu)$.
Between $\alpha$ and $\beta$ they are parabolas with 'acceleration' $1$;
the point of intersection determines a new exceptional slope $\gamma$. 
$\Delta_\gamma$ can be determined with the help of \lvdim.
By starting with two neighboring integers $\alpha=n$ and
$\beta=n+1$, this gives
a recursive description of $\epsilon$. It can be shown that the
rank $\rk_\gamma$ of the new exceptional bundle is given by
\eqn\newrank{
\rk_\gamma=\rk_\alpha \rk_\beta(3+\alpha-\beta).
}

Around each $\alpha=\epsilon({p\over 2^q})\in\CE$ there
exists a symmetric interval $I_\alpha$ defined by
$P(-|\alpha-\mu|)-\Delta_\alpha>\half$. In this interval
$\delta(\mu)$ can be written as
\eqn\deltaaltern{
\delta(\mu)=\delta_\alpha(\mu)=P(-|\alpha-\mu|)-\Delta_\alpha.
}
The intervals shrink for increasing $q$ and thus the function
$\delta(\mu)$ has a fractal structure near $\Delta=1/2$.

\ifig\deltaplot{The function $\delta(\mu)$}
{\epsfxsize4.0in\epsfbox{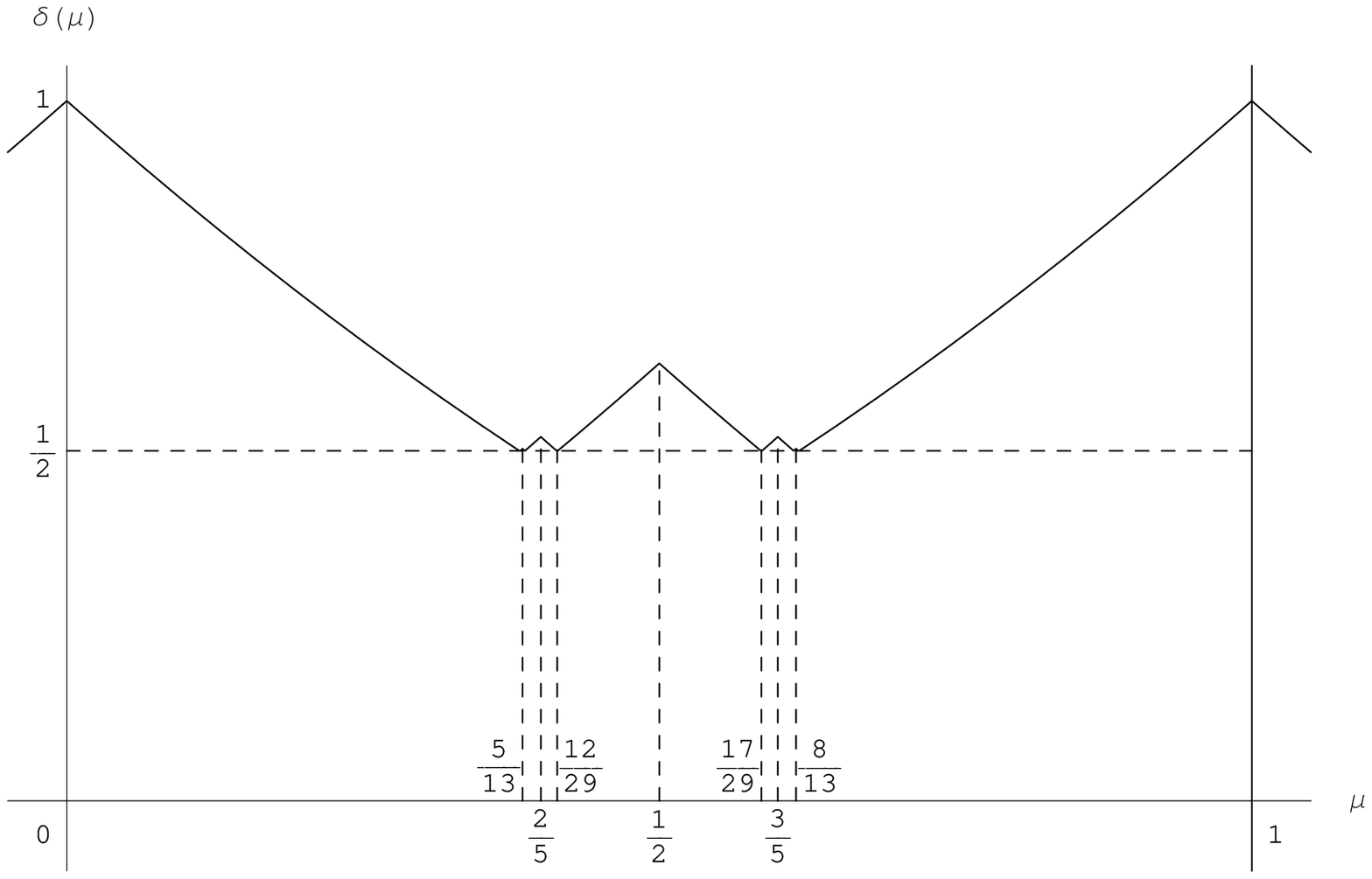}}

\subsec{The spectrum near the orbifold point}

As discussed in the appendix, the spectrum of BPS branes near the
orbifold point consists of the $\theta$-stable representations, where
$\theta$ are the Fayet-Iliopoulos terms up to a possible overall shift.
They are determined from the moduli \pistability. As we will describe 
below, the spectrum near the orbifold very strongly depends on the 
angle by which we approach this point.

Although the $\BZ_3$ quiver has $\BZ_3$ symmetry, the spectrum does not,
because of the non-zero Fayet-Iliopoulos terms.  If we come down
from the large volume limit along a particular path (we will generally 
consider the negative real axis as in section 2), we will obtain a particular
sign for these terms, say $\zeta_3>0$, $\zeta_2=0$ and $\zeta_1<0$.  This will
pick out one of the three sets of links as naturally zero; the reductions to
the Beilinson quiver which take this set non-zero will not have 
$\theta$-stable
representations.  Thus the basic picture of the spectrum on this line is the
stable representations of one of the Beilinson quivers.

Again, the basic criterion for a simple representation to exist is that
the expected dimension of moduli space satisfies $d\ge 0$. Since the 
expected dimension
formula was the same as that for large volume bundles, this leads to exactly
the same inequality \regularquad\ and exactly the same spectrum of exceptional
representations.  However there is an additional constraint, namely that
all $n_i\ge 0$ or else all $n_i\le 0$.  There is clearly no $\BZ_3$
quiver theory which can represent an object with some $n_i n_j < 0$; this 
would be a bound state of branes and antibranes at the orbifold point.

Although in general one can certainly obtain BPS branes as bound
states of branes and antibranes, a basic claim implicit in \dm\ and
supported by all subsequent work on this subject is that very near 
the orbifold point, the theories
obtained by the quotient prescription describe all branes
(i.e. describe all boundary states).  One does not need to take bound
states of branes and antibranes and would get a redundant description
of existing branes if one did (this is for branes at a single fixed
point; one can get new states by taking branes and antibranes at
different fixed points). Thus we 
consider this inequality to be a correct statement about the spectrum near 
the orbifold point.  In particular, states such as the D$2$ which would have 
gone massless there do not exist near the orbifold point.

Besides this additional inequality, the condition for stability is
different.  It depends on the particular line coming into the orbifold
point, which determines the Fayet-Iliopoulos terms in the quiver
theory, but nowhere near the orbifold point is it the same as the
large volume $\mu$-stability for all the D-branes.

One can easily check that neither the inequalities $n_i>0$ nor the
$\theta$-stability conditions are invariant under tensoring by line
bundles or equivalently $B\rightarrow B+1$, and it is easy to see that
the BPS spectrum is not either (e.g. by considering images of the
fractional branes).  Indeed, from the stringy point of view, such a
symmetry was not to be expected, but this is already a striking
difference from $\mu$-stability.

Let us discuss some aspects of the $\theta$-stability condition.
As explained in \pistability, an object with charge $n$ will decay into
products including a subobject of charge $n'$ on the line
\eqn\thetastability{
{\zeta\cdot n'\over e\cdot n'} = {\zeta\cdot n\over e\cdot n}
}
(where $e$ is the vector with components $e_i=1$). The relation between
the $\theta$'s we use to check $\theta$-stability and the physical FI terms 
is \pistability
\eqn\findtheta{\theta =\zeta-{\zeta\cdot n\over e\cdot n}e.} 
This relation can be derived by requiring a quasi-supersymmetric vacuum 
for BPS branes. 

In terms of the angle $\phi$ by which we approach the orbifold point
we have $\zeta_n = r \sin ({2\pi n\over 3}+\phi+\pi)$.
We have taken $\phi=\pi$, i.e. the negative real axis,
as our standard line to the large volume limit.
For orientation, we note that there are two other
lines $\phi=\pi+2\pi k/3$ which also lead to the large volume limit,
but with charges differing by a $\BZ_3$ monodromy.  Similarly the lines
$\phi=2\pi k/3$ are all lines to different copies of the conifold point.

We now give for every brane satisfying
$n_i n_j\ge 0$, a line leaving the orbifold point on which it
is most likely to be stable, using an argument similar
to the large volume argument for the necessary condition
for $\mu$-stability summarized in section 3.
Specifically, we want to find $\zeta$ such that 
for every $n'$ with $\theta\cdot n'<0$, we have $\chi(n',n)\le 0$,
which is compatible with the absence of a $\Hom(E',E)$.
If we take
$$\zeta = (\sgn n)\ \chi\cdot n$$
where $\chi$ is the matrix in \chiorbbasis, we have
$\theta\cdot n' = \chi(n',n) - \chi(n,n) e\cdot n'/e\cdot n$.
This works for a regular brane since it has $\chi(n,n)\le 0$.
An exceptional brane has $\chi(n,n)=1$, and since $e\cdot n'<e\cdot n$
for a subobject, and $\chi(n',n)\in\BZ$, this also works.

This is the analog of the necessary condition in large volume, but
we have not proven its sufficiency.  Nevertheless
it seems to work in examples, so we will go on to assume it is
sufficient in building our picture.

We next argue that each brane stable at large volume exists on a
wedge near the orbifold point $O$ containing this line.  
First, if a brane is stable at a point $z$ near $O$, it 
can not be stable at the point $-z$.  Thus, if we start from the
line and move either clockwise or counterclockwise near $O$,
we must encounter lines of marginal stability.  One can also see from
\thetastability\ and the definition of subobject that depending on
whether we move clockwise or anticlockwise,
the subobject that triggers the decay will be different.

One must look at the list of subobjects to find these lines.  It may well
be that as at large volume, the results will be summarized in a simple
condition, but we have not studied this systematically, rather we considered
examples.

The simplest example is that all branes with $n_3>0$ have $(0\ 0\ 1)$ as a 
subobject, corresponding to the need to have $\theta_3>0$ for a bound state.
Thus they will decay at the latest when $(\zeta_3 e - \zeta)\cdot n=0$, i.e. 
$\tan\phi = -(1+2n_1/n_2)/\sqrt{3}$. One can easily find examples in which
there is no previous decay and this result  implies that an infinite number
of marginal stability lines come into the orbifold point.  

Another possible subobject is $(0\ n_2\ n_3)$, which
would trigger a decay at $\tan\phi = (1+2n_3/n_2)/\sqrt{3}$. 
Notice that this subobject is not necessarily Schur (simple), but that if
so there will be a subsubobject which leads to an earlier decay.
There are examples where it is Schur, giving concrete examples of
decays not triggered by a fractional brane.

\newsec{The spectrum at general points}

In this section we will study the spectrum of $\Pi$-stable objects
along a line (the negative real axis)
connecting the large volume and orbifold points.
Besides getting a more detailed picture of how the BPS spectrum
varies, we will be testing whether $\Pi$-stability indeed describes
sensible physics.

An object $E$ is $\Pi$-stable if all of its subobjects $E'$ satisfy
\eqn\pistabilityeqn{
\grade(E') < \grade(E),
}
where the grade is defined in terms of the BPS central charge $Z(E)$ as
$\grade(E)={1\over\pi}\Im\log Z(E)$.

This presupposes that we know all objects and their subobjects, in other 
words we know the category of holomorphic objects. Let us first discuss 
marginal stability near the large volume limit, where it is plausible to 
assume that the category is still the category of coherent sheaves.

\subsec{Marginal stability near the large volume limit}

An interesting question is whether there are lines of marginal stability
arbitrarily close to the large volume limit. Using $\Pi$-stability we will
argue that this is indeed the case. To see this, we 
expand the asymptotic form of the central charges \zchargea\ in terms of 
$t=B+iJ$, for small differences in the phase (no grading) and we obtain
\eqn\lvpistability{
\left({J^2\over 2}+{B^2\over 2}-{1\over 8}\right )(\mu-\mu')-(\mu-B)
{\hbox {ch}_2'\over r'}+(\mu '-B){\hbox {ch}_2\over r}>0
}
as the modified stability condition. In the limit $J\rightarrow\infty$ 
this reduces to $\mu$-stability. If $\mu=\mu'$, the 
criterion is for $\mu>B$
\eqn\giesecker{
{\hbox {ch}_2\over r}>{\hbox {ch}_2'\over r'}
}
This is known as Gieseker stability in the math literature\foot{For
a recent appearance of Gieseker stability in the physics literature
see \mamost.}. Note however, that
for $\mu<B$ we get the opposite inequality. It would be interesting 
to know the significance of this.

We now construct a sequence of bundles stable in the large volume limit,
$E_n$, which by choosing sufficiently large $n$
will decay arbitrarily close to the large volume limit, according to
\lvpistability.  In fact, this will be true of any sequence such that
all $E_n$ have $\CO(1)$ as a subobject, all $\mu _n>1$, 
in the $n\rightarrow \infty$ limit $\mu _n\rightarrow 1$, and 
$ch_2(E_n)<C<<0$ for some $C$.

Now, the conditions for stable bundles are such that it is easy to find
sequences $F_n$ with the same properties except that we do not require
that $\CO(1)$ is a subobject.  
One can check that $\chi(F_n(-1),\CO(1))<0$ using \defchi, so
such bundles will always have
$\Ext^1(F_n(-1),\CO(1))\ne 0$.
Thus the extensions 
$0\mapr \CO(1) \mapr E_n \mapr F_n(-1) \mapr 0$
provide such a sequence.

\subsec{Comparison of the holomorphic categories}

We know that at large volume, the relevant objects are coherent sheaves on
$\P^2$, while near the orbifold point, the relevant objects are
representations of the $\BZ_3$ quiver.  As we explained, these categories are
extremely similar.  However, they are not literally the same.

We already made the main points in the previous discussion.
On the one hand, the map between Chern classes of sheaves
and the integers $n_i$ of the fractional brane basis does not preserve
positivity: some sheaves have charges $n_i\ge 0$
and can be represented by the Beilinson monad, while others have mixed
signs for the $n_i$ and cannot be so represented.
On the other hand, the orbifold category has at the
very least a three-fold degeneracy compared to the large volume category
coming from the $\BZ_3$ symmetry and the fact that we can set any of the
three links to zero to get a Beilinson quiver.
Thus neither category strictly contains the other, so we cannot simply
use the larger of the two along our trajectory.

The $\BZ_3$ multiplicity of the orbifold category has no obvious analog at
large volume (though see \dtoappear), so it is not too clear how to
base our discussion on the category of sheaves.
On the other hand, physically it is fairly clear
what we want to do to get orbifold states with mixed signs of $n_i$: we
want to allow bound states of fractional branes and their antibranes.
Although we made a point of saying that this was not necessary near the
orbifold point, it is a priori quite plausible that this could lead to
new BPS branes away from the orbifold point, and there are many
concrete examples such as the D$2$ which look just like this.

Such bound states are naturally described by complexes,
and thus the natural category to consider is that of
complexes made from the $\BZ_3$ quiver representations.  Beilinson's
theorem even provides a fairly explicit way for getting such complexes
associated to particular coherent sheaves.  The problem with this
however is that these complexes are not at all in one-to-one
correspondence with sheaves; rather, many complexes can have the same
sheaf as cohomology, and most of the physical properties of the brane
(stability, the moduli space, and so on) depend on which one we take.
Ideally, this would not be a problem as one would find that at most
one complex was stable at each point in moduli space, but this is not
a priori obvious.

A mathematical construction which gives us a unique representative for
each complex is to form the derived category from the original
category.  This construction is explained in
\refs{\gm,\kontsevich,\thomas,\sharpedc,\tomasiello} and in
\dtoappear\ which gives additional reasons this is 
useful in our string theory application.  Indeed, Beilinson's theorem
was originally phrased in this language: the derived categories formed
from the two categories we discussed are equivalent.  However, at
least at present, we do not know how to use the derived category in
discussions of stability, for reasons also explained in \dtoappear.

Having said this, we proceed to work with the category of
complexes of $\BZ_3$ quivers, and give examples of how some of the
decays which we know must take place on the way to the orbifold point
could work.

\subsec{Flow of gradings}

So far we have only discussed the comparison between the holomorphic objects 
in the large volume and orbifold regimes, but for the categories to agree,
the morphisms must also agree.  As is implicit in \kontsevich\ and as
discussed in \dtoappear, this will only
be true if we take into account a spectral flow of their gradings which is
determined by the flow of the brane central charges.

A simple example which is relevant for stability is the pair of branes
$\CO$ and $\CO(-3)$.  Both are stable at large volume and on the negative
real axis coming into the orbifold point, but on the opposite (conifold)
line $\CO(-3)$ will decay.  This is clear from its fractional brane charge
$(6\ 3\ 1)$, and in terms of $\theta$-stability happens because $(0\ 0\ 1)$,
which is $\CO$, is a subobject.

The confusing thing about this is that $H^0(\CO,\CO(-3)) = 0$, so this
subobject relation is certainly not true at large volume.

However, this comes about because in the large volume limit, 
as follows from Serre duality, one has 
\eqn\dimhtwo{
\dim H^2(\CO,\CO(-3)) = 1
}
and the flow of the gradings turns this into the $\Hom$ of the orbifold point.

A useful notation is to indicate the gradings of both branes and morphisms by
square brackets, so the homomorphism we described becomes
\eqn\derivedcat{
\dim \Hom(\CO,\CO(-3)[2]) = 1.
}

From section 2 one sees that as one goes from large volume to orbifold,
the two periods go around the origin on different sides.
This means that the relative grading is shifted by $2$, and the result
\derivedcat\ turns into
\eqn\derivedcat{
\dim \Hom(\CO,\CO(-3)) = 1
}
for the natural (zero grading) objects at the orbifold point.

\subsec{Another technical result on $\Pi$-stability}

In practice it is much easier to see whether there are homomorphisms
between objects, than to see whether they are injective.  
For $\mu$-stability
one circumvents this difficulty by arguing that if there exists a
$\phi\in\Hom(E',E)$ from a {\it stable}
object with $\mu(E')\ge \mu(E)$, then $E$ will be destabilized,
either by $E'$ or by the preimage of $\phi$, which will be
an object $E''$ with an exact sequence
\eqn\exactseq{
0\mapr E''' \mapr E' \mapr E'' \mapr 0.
}
Because $E'$ is stable, $\mu(E'')>\mu(E')$.  
This follows from $\mu(E''')<\mu(E')$, $r'''+r''=r'$ and $c_1'''+c_1''=c_1'$.

The same is true of $\Pi$-stability, if the differences between the
gradings of any two objects are less than $1$.  This follows from the
same argument but deriving the relation $\phi(E') < \phi(E'')$ from
$Z'=Z''+Z'''$ and the convexity of $\phi=\Im \log Z$ under this assumption.
It may be true more generally, but we have no argument for this.

\subsec{The D2-brane is born, and other examples}

In this section we want to motivate the line of marginal stability for 
the D2-brane. The orbifold charges of the D2-brane are $(1\ 0\ -1)$. This
brane can obviously not exist at the orbifold point, but it exists in the 
large volume. Looking at the evolution of the periods of the fractional 
branes in \periodplot\ one finds that at a point $P$ on the negative 
$\xi$-axis the periods $\Pi_1$ and $\Pi_3$ become antiparallel.  This is
the obvious proposal for the location of
a line of marginal stability for the D2-brane, at which it
decays into $(1\ 0\ 0)$ and $(0\ 0\ -1)$, in other words
$\bar\CO(-1)$ and $\CO$.

For $\Pi$-stability to describe this process, it must be that one
of these objects is a subobject of the D$2$.  We can describe the
D$2$ by a sheaf supported on a two-cycle, $\CO_\Sigma$, defined by
the exact sequence
$$
0 \mapr \CO_i \mapr \CO \mapr^\phi \CO_\Sigma \mapr 0
$$
where $\CO_i$ is the ideal sheaf of $\Sigma$, i.e. functions vanishing
on this divisor.  Clearly the map $\phi$ is our candidate homomorphism.

An interesting point about this is that the large volume
gradings between branes of complex dimensions differing by an 
odd integer are naturally half integral 
(from ${1\over\pi}\Im\log (B+iV)^p$).
We assign $\phi$ in $\Hom(\CO[0],\CO_\Sigma[\half])$, which will
flow to zero degree at the point $P$.

One also might not have thought that $\phi$ was injective from the
large volume definition.  As all of the 2B's will decay into combinations
of 4B's, this is a very general problem.  The result of the previous
section tells us that we could still have a decay, but this would go into
an object different from $\CO$
(the $E'''$ above), which seems rather implausible without
further evidence.

The $\Hom$ is injective on global sections, so perhaps this is the correct
definition.
We can also see that there should be an injective $\Hom$ (and get some
ideas for a more general definition of the category) by constructing a
complex of $\BZ_3$ quivers which reproduces the 2B.
This can be done in various ways by taking two trivial (single node)
quivers for the two constituents and adding brane-antibrane pairs.
The simplest way is to take $(1\ 0\ 1)$ and $(2\ 0\ 0)$ to represent
the antibrane and brane respectively, use the third link of the $\BZ_3$
quiver (the one which was zero in the Beilinson quiver) to make
a simple object out of $(1\ 0\ 1)$, and then postulate two ``tachyon''
links of opposite orientations between brane and antibrane which are turned
on to make the bound state.  The moduli space of the resulting complex
is indeed $\P^2$ as is correct for the 2B, and one sees that there is
an injective $\Hom$ from $(0\ 0\ 1)$ into this complex.

Although this construction is not unique, it shows that there is a
definition of the 2B consistent with the general picture we are suggesting.

Another interesting example is the bundle
$\Omega^1(2)=T(-1)$, an exceptional rank 2 bundle. 
Its only subsheaves are $\Omega^1(1-n)$ and $\CO(-n)$ for $n\ge 0$.
From the Euler sequence one can determine its charges to be
$(1\ 0\ -3)$, and according to $\Pi$-stability it is
destabilized by $\CO$ at the same point $P$ as the D2-brane,
where it decays into $3\ \CO+\bar\CO(-1)$.

Since there is a $\BZ_2$ symmetry of the moduli space, the reflection on
the real $\xi$-axis, which acts on the spectrum by exchanging $n_1$ and
$n_2$ ($\mu\rightarrow -1-\mu$), the brane $(1\ 0\ -3)$, i.e. $\Omega^1$ 
has the same line of marginal stability, but there is no subsheaf 
destabilizing it.  By analogy to the decay above,
the obvious candidate for the subobject triggering the decay is $\bar\CO$. 

In fact, a homomorphism in $H^1(\CO,\Omega^1)$ can explain this decay,
taking into account the flow of gradings.  These must exist because
$\chi(\CO,\Omega^1)=-2$.

\subsec{The conifold point and the general picture}

The conifold point would be a particularly interesting test of the
general formalism as the issue of how to treat large variations of
the gradings comes to the fore here.  This is simply because a closed
loop around the conifold point increases the grading of $\CO$ (the
brane which becomes massless there) by two.

We do not have a solvable perturbative string theory realization of
this region of the moduli space, but the basic physics is quite clear
from the known relation to Seiberg-Witten theory \geometric\ %
and the results so far.  On the large volume side of the conifold point,
there is no reason to expect the spectrum to be drastically different
from the large volume spectrum.  On the other hand, we know that if we
encircle it, we would produce bound states with arbitrarily large D$4$
number which do not exist in the large volume spectrum; therefore there
must be lines of marginal stability through the conifold point on which
all of these decay.  The vanishing of the D$4$ central charge there means
that any state containing D$4$ charge (in some basis) will indeed see
candidate marginal stability lines on which it will decay to the D$4$
and other products.  

Since we have two other charges
in the picture, not just the one of gauge theory, 
we need additional information to decide
which states exist on the other side.  We will do this by outlining a 
global picture of the marginal stability lines we have already found.

Now, making a truly global picture would require a good description
of the ``Teichm\"uller space,'' the universal cover of the thrice-punctured
moduli space, or at least some cover on which periods are single-valued.
Since this is a one dimensional moduli space with a natural metric of
negative curvature, one might imagine that it could be described as some
quotient of the upper half plane.

Such a description is not known.  One serious complication compared
to the
familiar examples such as gauge theory where this is possible, is that
the period map, which in this example takes the Teichm\"uller space into
$\BC^2$ (this is local mirror symmetry so $\Pi_{D0}=1$), behaves badly
on the boundary (it seems to map it into a fractal in $\BC^2$).

Not having this, we restrict attention to the triple cover of moduli
space near one orbifold point $O$, parameterized as in the introduction.
Near the orbifold point, this cover will have
three conifold points C$_i$ where the $i$th fractional brane $B_i$
becomes massless. There are also three large volume points LV$_i$ opposite 
the conifold points. The large volume point that we normally consider
is LV$_3$.  The monodromies associated with these points are described
by branch cuts in the periods which leave this region, and the full
Teichm\"uller space will be covered by copies of this region in which
the role of the fractional branes is played by
monodromy images of our three fractional branes.\foot{
This part of the story is related to the theory of helices. \seidthomas}

\ifig\marglines{The structure of the Teichm\"uller space}
{\epsfxsize2.5in\epsfbox{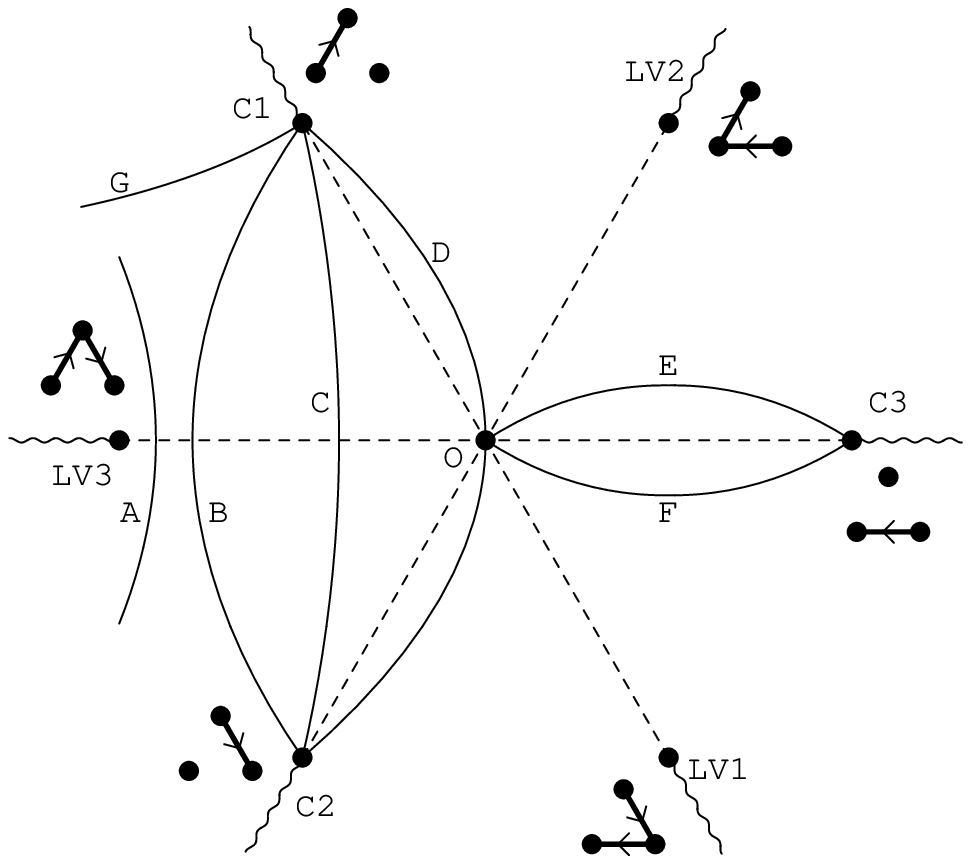}}

We can now trace the marginal stability lines we found near the
orbifold point and intersecting the negative real axis out into this
region.  As we discussed,
the natural endpoint for a line describing a decay into a
fractional brane $B_i$ is the conifold point C$_i$.
We give a map in \marglines\ with various lines which we proceed
to explain.

We can distinguish two basic types of marginal stability lines at this
point.  On the one hand, we found lines coming out of the orbifold point
with decays triggered by fractional branes.  The simplest picture is
that these head along a (fairly direct) path to the appropriate conifold
point, as in lines D, E and F on the figure.  
We will refer to these as ``CO'' lines.  .

We also found decay lines for branes such as the D2 which do not exist
near the orbifold point.  Now the D2 and the similar objects
$(n_2\ 0\ n_1)$ will decay into a fractional brane and a different
fractional anti-brane.  The simplest picture is that such a line connects
the two associated conifold points, as do lines B and C on the figure.
We will refer to this as a ``CC'' line.

In general, both types of line will also exist for decays into
other exceptional branes, and should lead to the conifold points in
other regions of the Teichm\"uller space, as do lines A and G.
There might also be decays
purely into regular branes.  Since monodromies do not in general
preserve the dimension formula \lvdimformula, one might wonder if
these could also be associated with monodromy images of exceptional
branes.  The strong decoupling hypothesis \dtoappear\ 
requires all such monodromies which change the dimension of
moduli space of the holomorphic objects
to cross lines of marginal stability, and would disallow this;
then these lines would presumably be ``OO'' lines
between different orbifold points.

Note that just because a marginal stability line crosses the negative
real axis does not imply that it is a CC line.  For example, by tensoring
the bundles we discussed as examples of decays near the large volume limit
with appropriate line bundles, one can get examples of decays on the negative
real axis of objects
with all $n_i$ positive.  These must exist near the orbifold point so
these are in fact CO lines.

There are a few more general considerations we can use to constrain the
picture.  First,
there appear to be no marginal stability lines from a singularity to itself.
For the orbifold this is because (as discussed earlier) the two 
marginal stability lines for a given object are triggered by different
subobjects, while for the conifold this should follow because we
do not expect an object to decay both into a brane and the same antibrane.

Two marginal stability lines involving the same object cannot cross in
this problem, except at orbifold points, because such a crossing
requires at least three periods to align.  In particular the CO lines
triggered by the same fractional brane cannot cross, which
makes the hypothesis that they take fairly simple paths well-motivated.

The symmetry of the problem about the negative real axis, and other
symmetries, should be useful.  We leave the problem of finding the
complete picture however to future work.

As for the question of how small the spectrum can get, these
considerations suggest that the smallest spectrum is already attained 
near the orbifold point and near its line to a conifold point.
The general character of the spectrum here is probably illustrated
by the decay to $(0\ 0\ 1)$ which as we discussed takes place on
a line with angle
$\tan\phi = -(1+2n_1/n_2)/\sqrt{3}$.
If we start on the line to LV$_3$ and decrease $\phi$ to $2\pi/3$,
branes with $n_1<n_2$ will decay before this line,
while branes with $n_1>n_2$ will have marginal stability lines such as D
(assuming they decay into $\CO$), and survive on this line.

Going farther will produce further decays of this type, and in this
sense the spectrum can become arbitrarily small.  On the other hand,
if we go too far, we will pick up objects naturally associated with
LV$_2$, and the spectrum will grow again. 

If we take the line $\phi=2\pi/3$ as representative, the overall
picture is one in which almost all bound states with the D$4$ do decay
as we go around the conifold, but along any given line, a large subset
(with three adjustable charges satisfying inequalities) do exist.

We finally comment on the role of the D$0$-brane, which naively one
might think would play a central role in the story.  Although the D$0$
appears to
exist as a BPS brane everywhere except at the orbifold point itself
in this model, this is to some extent a simplification of local mirror
symmetry (its conjugate, the D$6$, has infinite tension in this limit).
The D$0$ is clearly not the simplest object near the orbifold point and
is not naturally a subobject of other branes at large volume either,
since one does not have natural maps to sheaves of higher dimension.
Thus the results here support the idea that it (and all non-exceptional
branes) is best thought of as
a bound state of simpler objects, the exceptional branes. 

\newsec{Bound states on the quintic}

The approach we discussed can surely be extended to other orbifold
singularities and noncompact CY's (some relevant mathematical work is \naka).
But can these ideas describe branes and stability on compact CY's?

A natural way to try to describe bundles on varieties in projective
space is by constructing bundles on projective space and restricting
them to the subvariety.  Of course not all bundles can be so obtained,
but this is already a much more general class than has been available
in any of the constructions commonly used by physicists (such as \fmw).

Although the complete analysis of stable bundles is not known for
$\P^d$ with $d>2$, many of the ingredients we described are known.
In particular, Beilinson's construction of holomorphic bundles
generalizes directly to $\P^d$.  Given a bundle $E$ on $P^d$,
the data $H^p(E(-q))$ and natural maps between these spaces defines
a spectral sequence whose cohomology is $E$, and in favorable cases
this again reduces to a complex whose cohomology is $E$.
The complex is the obvious generalization of \monad:
\eqn\monadN{
0\rightarrow \BC^{n_d} \otimes \CO(-1)\mapr^{\hat X^{d-1}} 
\BC^{n_{d-1}} \otimes \Omega^{d-1}(d-1) \mapr^{\hat X^{d-2}} 
\ldots
\mapr
\BC^{n_2} \otimes \Omega^1(1) \mapr^{\hat X^0 } 
\BC^{n_1} \otimes \CO
 \rightarrow 0
}
with $X^n_{[i} X^{n-1}_{j]} = 0$.
In quiver terms, we have $d+1$ nodes in a line, with $d+1$ arrows
between successive nodes.

It turns out that there is a very direct analog of this structure implicit
in the results of \bdlr\ describing rational B branes in the $(3)^5$ Gepner
model \rs.  These branes are labeled by quantum numbers $0\le L_i\le
1$ in each minimal model factor and an overall $0\le M<10$, which is
even for branes and odd for antibranes.

Since the elementary bundles $\Omega^n(n)$ of this construction are
all rigid, their natural counterparts are the $L=0$ branes, of which
we take the five branes $M=2i$.  We thus consider an $\CN=1$ quiver gauge
theory defined using any number $n_i$ of each of these branes, and
chiral multiplets containing massless fermions stretched between pairs
of branes.  This spectrum can be read off from the results of
\refs{\rs,\bdlr}: it is a $\prod U(n_i)$ theory with five multiplets
$X^a_{i,i+1}$ in each $(n_i,\bar n_{i+1})$ and ten multiplets 
$Y^{[ab]}_{i,i-2}$ in each $(n_i,\bar n_{i-2})$. (Here $1\le a,b\le 5$
are global quantum numbers).

We immediately notice that the links $X$ are in exactly the same
relation to the Beilinson complex for $\P^4$ as the links of the
$\BZ_3$ orbifold model were to the monad for $\P^2$, if we set $X_{5,1}=0$.
The relations suggest the following ansatz for the superpotential:\foot{
This term in the superpotential can also be seen in CFT. \brs}
$$
W=\sum_{i,a,b} \tr X_{i,i+1}^a X_{i+1,i+2}^b Y^{ab}_{i+2,i},
$$
F flat configurations of this gauge theory with $Y=X_{5,1}=0$ are exactly
representations of the Beilinson quiver.

Furthermore, according to CFT, the bosons in the $X$ multiplets are 
tachyonic, so by condensing them one can find bound states of branes.
The natural conjecture is that these bound states are the Gepner
point images of a large set of bundles on the large volume quintic,
those obtained as the cohomology of the complex \monadN.

In particular, this identification predicts that the bundles corresponding
to $L=0$ branes are the bundles $\Omega^n(n)$.  One of the $L=0$ branes was
found in \bdlr\ to be the D$6$-brane $\CO$; it is easy to check that its images
under the $\BZ_5$ monodromy of the Gepner point
have exactly the predicted Chern classes.

We have furthermore identified some of the simpler bound states with
$L>0$ branes.  The simplest case is the bound state $(1\ 1\ 0\ 0\ 0)$
which according to the quiver gauge theory has moduli space dimension
$4$.  This is exactly the charge of a rational boundary state with
$\sum L_i=1$ and indeed the number of marginal operators for this
boundary state is $4$.  We made the same check for the bound state
$(1\ 2\ 1\ 0\ 0)$ which is the charge of a $\sum L_i=2$ brane, and
match the correct number of marginal operators $11$.

These results begin to answer one of the main questions raised 
in \bdlr\ and related work -- the large volume
identification of the B rational boundary states.  Of course for these
to actually exist as BPS states
at large volume, they must be stable bundles.  There
is a further correspondence
in that the brane predicted to decay in \dreview\
is in fact a bound state which would not exist at large volume (it
uses the link $X_{5,1}$ which is not present in the Beilinson complex).

\newsec{Conclusions}

We studied the spectrum of BPS branes on the non-compact Calabi-Yau
space ${\cal O}_{\P^2}(-3)$, and gave detailed results for the spectrum near
the large volume limit and near the orbifold point, and a qualitative
picture elsewhere.

We see this work as shedding light on at least two aspects of string
physics on Calabi-Yaus: besides describing a highly non-trivial
example of an $\CN=2$, $d=4$ theory, we can also interpret our results
as providing a picture of the set of possible gauge bundles which
could be used in type \I\ and heterotic string compactifications, and
suggesting new ways to analyze this aspect of $\CN=1$
compactification.  Let us discuss these points in turn.

There have been many studies of the spectrum of BPS states in $\CN=2$
theories and marginal stability, mostly on systems with a fairly
simple spectrum, supersymmetric gauge theories and the corresponding
limits of string theory compactification.  In these theories the weak
coupling spectrum is that of dyons with highly constrained magnetic
charge but fairly general electric charge.  One or a few lines of
marginal stability separate the weak and strong coupling regimes, and
in the latter the spectrum is drastically reduced, to a finite number
of states, all of which become massless at singular points in moduli
space.

The picture here is broadly similar but with noteworthy differences.
The spectrum in the large volume limit is roughly characterized as
those charge vectors satisfying a quadratic inequality guaranteeing
that the moduli space dimension is non-negative, but there is a subtle
fine structure of the boundary of this region in which certain
rigid exceptional bundles exist and destabilize other regular
bundles.  As one moves to the orbifold limit, additional inequalities
come into play and reduce the spectrum.  The spectrum becomes
even smaller between the orbifold and conifold points but is
still infinite; it appears to never reduce only to states which can
become massless.

Our approach was to use a generalization of the idea of stability
which governs the existence of solutions of the hermitian Yang-Mills
equations and thus would determine the spectrum of BPS branes in the
large volume limit.  As discussed in \pistability\ and \dtoappear,
this approach divides the problem into that of studying the
category of holomorphic brane configurations and then finding the
stable objects in this category.  The approach is proven near the
orbifold point and is very well motivated near the large volume
limit, where it only assumes that mirror symmetry works.  

Quite strikingly, the category of holomorphic brane configurations is
almost the same in these two limits: the orbifold theory essentially
contains Beilinson's general construction of holomorphic bundles on
$\P^2$.  We consider this evidence for a strong form of the ``decoupling
conjecture'' of \bdlr, that the holomorphic category is largely independent
of K\"ahler moduli.  This similarity allows us to get at least partial
results on lines connecting these two limits.  It suggests that direct
application of mathematical results on this category and the related
derived category of sheaves such as those of \refs{\seidthomas,\horja}
will allow pushing this analysis to the entire moduli space.  It would
be particularly interesting to study the conifold point using these ideas.

Of the various alternate approaches one might take to this problem, we
should mention that of \stringnet, in which special Lagrangian cycles
on the mirror are reduced to string networks in F theory.
At present 
this would seem to be the most promising way to get results from the A
picture, and it will be interesting to see if this approach
can reproduce the very detailed structure we saw here.

We discussed the classical theory of these branes, but let us make a
few remarks about the quantum BPS particles in \IIa\ string theory on
this background.  These will be vacua of the quantum mechanical
theories obtained by dimensional reductions of the brane world-volume
theories we derived.  Typically, such vacua correspond to cohomology
classes of the moduli space.  One very general result which follows
from this identification is that typically (though not always),
crossing a line of marginal stability completely eliminates the moduli
space, which means that all the quantum branes of that charge will
decay on that line.  More specific results might be obtained by using
our explicit construction of the classical moduli space as a
symplectic quotient of the moduli space of quiver representations.
Many partial results exist on this spectrum
for large volume $\P^2$, in particular \refs{\vafawit,\katzklemmvafa}.
These results should also further the study of BPS algebras \hm.

We turn to the $3+1$ case.  The study of gauge bundles on Calabi-Yaus
has long been a central part of superstring physics.  At least
conceptually, the direct way to study $\CN=1$ compactification and
$\CN=1$ duality would be to have a list of Calabi-Yaus (quite a good
list of these is known, and this was very helpful in studying $\CN=2$
duality) and then a list of all sheaves on them.  Not only do we not
have such a list, there is no picture we are aware of in the physics
literature which gives any overall sense of the possibilities.  Even
very general questions such as in what range of Chern classes do stable
sheaves generically exist or generically do not exist go unaddressed.

We hope that the present work will serve to illustrate what such a
picture might look like, and believe that the ideas and techniques we
discussed will apply to branes in very general Calabi-Yau
compactifications.  We even gave the beginnings of this for the quintic.
A detailed study of this and other examples and the precise relation
to large volume bundles is under way; recently in \ddm\ we have found
how to generalize the results of section 7 to a large subset of the Gepner
models.

The most direct way to use such results to study $\CN=1$
compactifications is to choose a brane configuration and do an
orientifold projection, leading to perturbative type \I\ models
satisfying the known constraints such as tadpole cancellation.
Indeed, the explicitness of the
supersymmetric gauge theory description in the $\BC^3/\BZ_3$ and
$T^6/\BZ_3$ examples has allowed much work to be done, starting with
\abpss, including \morezthree\ and many others.

Although from this point of view one might consider the
present work as one more in a long list, the point at which we feel
we have made significant progress is to provide a clear algebraic geometric
interpretation of the brane configurations as specific bundles on the
resolved orbifold singularity, and explain how this geometric
interpretation and other relevant mathematics can be used to organize
and solve the problem of classifying supersymmetric vacua.
We furthermore feel that having the geometric interpretation will be a crucial
element in the study of superstring dualities involving these examples.

Clearly having such explicit results for compact Calabi-Yaus would
allow a much more complete study of $\CN=1$ compactification.  
This might seem rather optimistic for a problem with such a long
history; time will tell.  Let us conclude by listing the elements of the
emerging picture of D-branes on Calabi-Yaus which may provide
advantages over other approaches.

\item{$\bullet$}
Decoupling of complex and K\"ahler structure is a significant a priori
simplification and appears to allow detailed comparison between
constructions in different limits of K\"ahler moduli space.

\item{$\bullet$}
By starting with the classical limit, we can
work with configurations which do not satisfy tadpole or
anomaly cancellation constraints.  This allows building bundles out
of simpler constituents.

\item{$\bullet$}
Our results strongly suggest that
branes in the stringy regime are best thought of
as bound states of a finite set of rigid 
exceptional branes.  
Indeed, identifying an appropriate basis of bundles
in terms of which the others are simple bound states appears to be a
central theme in the mathematics (e.g. see \kapranov).
For Fano varieties (such as $\P^2$) one has a precise idea of what
one is looking for, an exceptional collection.  
Such collections do not exist for Calabi-Yaus, but we believe that
as in our quintic example,
the subset of rigid rational boundary states will provide
the appropriate generalization of this idea.

Finally, let us state
what we feel is the broadest lesson of this work.
In seeing how the
various elements of the problem of classifying bundles on $\P^2$
translate into the language of supersymmetric field theory, we realize
that doing this translation is actually much easier than classifying
the bundles.  
In the case of flat space,
where the translation is provided by the ADHM construction,
this has been appreciated by physicists for some time now \smallinst.

Although classification on a Calabi-Yau is much harder, this does not
mean that the translation need be so much harder.
The hard parts of the classification, determining holomorphic moduli
spaces and stability, get translated into aspects of the problem of
finding supersymmetric vacua of the resulting theories, and the
greater difficulty of solving them is essentially for the usual reasons
that these problems can be hard in $\CN=1$ supersymmetric theory.

If such translations are generally possible, a part of the problem of
string compactification that many physicists (at least the authors)
have found rather mysterious and intimidating, will turn out to
be not so different from the more familiar parts.

\medskip

We would like to thank  A. Beilinson, I. Brunner, M. Cvetic,
P. Deligne, D.-E. Diaconescu, A. Klemm, M. Mari\~no, G. Moore,
D. Morrison, T. Pantev, V. Schomerus,
R. Thomas, C. Vafa, E. Witten and S.-T. Yau for helpful discussions,
communications and comments.  We especially thank D.-E. Diaconescu
for discussions and ongoing collaboration on the results in section 7.

This research was supported in part by DOE grant DE-FG02-96ER40959.

\appendix{A}{Quivers and their representations}

The theory of quivers is a framework encompassing a wide
variety of problems in linear algebra and representation theory; a couple
of nice references on the subject are \refs{\benson,\kraft}. Some relations 
between quiver theory and gauge theories have been discussed recently in 
\yang. 

A {\it quiver} $Q$ is a directed graph; it
can be defined by a set of vertices $V$, a set of arrows $A$
connecting vertices, and two functions $i$ and $t$ from $A$ to $V$
specifying the initial and final vertex for each arrow.

An {\it path} in the quiver has the obvious definition (the arrows
must preserve orientation).  The {\it path algebra} is the algebra 
generated by the paths, with multiplication given by concatenation.
Perhaps the simplest mathematical definition
of ``quiver theory'' is that it is the representation theory of such
algebras.

\subsec {Gauge theory and representations of quivers}

Associated to a quiver $Q$ there are a set of gauge theories $\CQ(\vec n)$
with gauge group $G=\prod_{v\in V} U(n_v)$ and bifundamental matter.
We will consider $\CN=1$, $d=4$ supersymmetric theories, with a chiral
multiplet $x_a$ associated to each arrow $a\in A$, in the $(n_{ta},\bar 
n_{ia})$ representation of $G$. 

A point in the configuration space of this theory then corresponds to
a {\it representation} $R$ of the quiver $Q$.  This is a collection of finite
dimensional vector spaces $R(v)$, one for each vertex $v\in V$,
and a collection of linear maps between these spaces $X_R(a)$, 
one for each arrow $a\in A$.  
The vector $\vec n$ with components $n_v=\dim R(v)$ is referred to as
the {\it dimension vector} $\dim R$ of the representation.

In supersymmetric gauge theory one also makes use of the complexified
gauge group $CG$ which would be $\prod_{v \in V} GL(\dim R(v))$.
It turns out to be
useful to phrase the relation of gauge equivalence between configurations
in a way that does not require inverting group elements.  
Thus we write a ``gauge transformation'' $\phi$ as a set of linear
transformations $\phi(v):R'(v)\rightarrow R(v)$ for which
\eqn\quiverisom{
X_R(a) \phi(ta) = \phi(ia) X_{R'}(a)
}
for all $a$.  If $\phi\in U(R)$ this is a true gauge equivalence.

This can be further generalized to a {\it homomorphism} of representations
$\phi:R\rightarrow S$: this is a set of linear maps 
\eqn\quiverhoma{
\phi(v):R(v)\rightarrow S(v)
}
satisfying
\eqn\quiverhomb{
X_R(a) \phi(ta) = \phi(ia) X_S(a).
}
Since we do not require invertibility, these form a group under addition,
called $\Hom(R,S)$.
If $R\cong S$ this group is the {\it endomorphism} group $\End R$.
Invertible (in the usual matrix sense)
endomorphisms are elements of $CG$ and thus the dimension of the
endomorphism group is also the dimension of the
the subgroup of the gauge group left unbroken by the configuration $X_R$.  
If there is an injective homomorphism from $R$ to $S$ we say that $R$ is a
{\it subrepresentation} of $S$. When the generic representation of 
dimension vector $\alpha $ has a subrepresentation with dimension vector 
$\beta $ we say that $\beta $ is a {\it subvector} of $\alpha$.

We will say that a representation is {\it indecomposable} if it can not be
written as a direct sum of smaller representations, while we say that
a representation $R$ is {\it Schur} (or simple), if $\End R
\cong \BC$.  Note that unlike the more familiar theories
of finite groups and compact Lie groups, we need to distinguish these 
notions: although a Schur representation is indecomposible, the converse
is not always true.  The Schur representations
are of particular relevance for us, since the breaking of the gauge 
group down to $U(1)$ signals the existence of a bound state.

We note in passing that the quiver representations are objects
in an {\it abelian category} with the homomorphisms as just defined.
Many of the further definitions and theorems are usually discussed in
 this more general context.

\subsec{Supersymmetric vacua of gauge theory}

In the gauge theory application, a primary problem is to find the
space of vacua of the theory.  In the case of supersymmetric classical
vacua of $\CN=1$ gauge theory, this means gauge equivalence classes of
configurations which satisfy
the ``D and F flatness conditions.''  We require two further pieces of
information about the theory to formulate these.  

The first is the
superpotential $W$, a holomorphic gauge invariant function of the chiral
fields.  The F-flatness conditions are then $\p W/\p x_a=0$ for all $a$.
Let us call the variety defined by these constraints the ``F-reduced
configuration space.''

The second is qualitative information about the kinetic term; we will
assume that it is non-singular.
In fact we will choose $K=\sum_a \tr x_a x_a^+$, but the qualitative
features of the result do not depend on this choice.
The D-flatness conditions then depend on $|V|-1$ real parameters
(the ``Fayet-Iliopoulos terms'')
and are
\eqn\dflatness{ 
\sum_{ha=v} X_a^+ X_a - \sum_{ta=v} X_a X^+_a = \zeta_v.
}
As is well-known these constraints modulo gauge equivalence
are an instance of the symplectic quotient construction.
Thus the moduli space of supersymmetric vacua is
the symplectic quotient of the F-reduced configuration space by $G$.

There is a further relation between this quotient and
a different definition of quotient provided by geometric invariant theory.
(This is fairly well known to physicists in the case $\zeta=0$,
e.g. see \luta). The GIT quotient is defined in terms of the set of
orbits of the complexified gauge group $CG$,
acting as $x\rightarrow g^{-1} x g$.  This is a space which
looks roughly like the symplectic quotient -- in particular, it has the
same complex dimension.  Furthermore, each orbit contains at most one
solution of the D-flatness conditions (for some FI terms).
However it turns out to be larger -- some of the
orbits do not contain any solution of the D-flatness conditions.

Very generally, geometric invariant theory provides a concept of
{\it stable} orbits which are precisely those which correspond to points
in the symplectic quotient, i.e. those containing a solution of
the D-flatness condition.  The basic idea will be that stable orbits
are those which are closed under a suitably extended form of the group action.
This means that if we try to find the solution by
minimizing a potential (the usual $D^2$, say) over the orbit, the minimum
must be attained.  Conversely, orbits which are not closed will not contain
this minimum.

In the context of quivers (or abelian categories), 
an appropriate definition of stability is 
{\it $\theta$-stability}, formulated by King \king.
Let $\theta$ be a vector indexed by $V$; then $R$ is 
{\it $\theta$-semistable} if $\theta\cdot\dim R=0$ and every subobject
$R'$ (i.e. one for which there exists an injective
homomorphism $\phi:R'\rightarrow R$) satisfies 
$\theta\cdot\dim R'\ge 0$.
Furthermore, $R$ is {\it $\theta$-stable} if the only subobjects with
$\theta\cdot\dim R'=0$ are $R$ and $0$.

The theorem is then that $R$ is a solution
of the D-flatness conditions for $\zeta$
precisely when it is a direct sum of $\theta$-stable representations
with $\theta=\zeta$.

This is a useful criterion because it translates the problem of solving
the D-flatness conditions into a purely algebraic question, that of
classifying possible subobjects, which can be studied inductively.
Furthermore it tells us how the existence of solutions depends on the FI
terms, as we discuss in examples in the main text.

We now consider the combined problem of D and F-flatness conditions.
Although a theory with general superpotential does not correspond to
a problem studied in quiver theory, superpotentials which can be
written as a single trace produce {\it quivers with relations}.
Each of the matrix equations 
$${\p W\over\p X^a} = 0$$
can be thought of as defining a weighted sum over oriented paths
in the quiver; equating the sum to zero
defines a relation in the path algebra.

The example of immediate interest for us is the superpotential
\eqn\quiversuperpot{
W = \epsilon_{ijk} \tr X^i_{1,2} X^j_{2,3} X^k_{3,1}
}
for which the F-flatness conditions $W'=0$ are equivalent to requiring
a set of commutation relations
\eqn\quiverfflat{
X^i_{n,n+1} X^j_{n+1,n+2} = 
X^j_{n,n+1} X^i_{n+1,n+2}.
}

As pointed out in \king, the discussion of $\theta$-stability works
the same way in the presence of relations.  This is because
if an object satisfies the relations, its subobjects
will also satisfy them.

\subsec{Roots of quivers}

The dimension vectors of representations of a quiver with $n$ nodes live in 
$\BN^n$. We introduce now many concepts similar to those of the theory of Lie 
algebras.  In this subsection we consider only quivers without relations.

Given the dimension vectors $\alpha $ and $\beta $ of two 
representations, we can define the {\it Euler form}
\eqn\eulerringel{
<\alpha, \beta>=\sum _v\alpha _v \beta _v-\sum _a \alpha _{ia}\beta _{ta}
}
and the {\it Cartan matrix}
of a quiver as the symmetrization of the Euler form
\eqn\eulercartan{
(\alpha,\beta)=\ \alpha ^T\cdot C\cdot \beta =\ <\alpha,\beta>+<\beta, 
\alpha>
}

The expected dimension of the moduli space of a quiver representation
(and thus of the gauge theory) is
\eqn\quiverexpdim{
d(\alpha) = 1 -{1\over 2}\alpha ^T\cdot C\cdot \alpha 
}
If $d\ge 0$, we expect to find such objects, while if $d<0$
we expect not to.  These expectations are made precise
in a theorem of Kac \kac, which we proceed to quote.

For each node of the quiver, consider the n-vector $e_i=(0,0,\dots,1,\dots,0)$
with the 1 in the $i$'th position. Let us call these $n$-vectors the 
{\it simple roots} of the quiver, and denote the set of them by $\Pi$. 
For each simple root we can introduce a Weyl reflection acting on an 
arbitrary vector $\alpha$:
\eqn\weyl{
r_i(\alpha)=\alpha -2{(\alpha,e_i)\over (e_i,e_i)}e_i
}

The {\it fundamental region} is defined as the set of vectors $\alpha$ 
satisfying $r_i(\alpha)\geq \alpha $ for all $i$, and whose
{\it support} (i.e. the subset of nodes for which $\alpha_i\ne 0$)
is connected.

The group ${\cal W}$ generated by the Weyl reflections is called the 
{\it Weyl group}
of the quiver. The {\it real roots} of a quiver are defined as the set 
of vectors obtained by the action of the Weyl group on the simple roots.
\eqn\rroots{
\Delta ^{re}={\cal W}(\Pi)
}
The {\it imaginary roots} of the quiver are the set of vectors obtained by 
the action of the Weyl group on the vectors of the fundamental region.
\eqn\iroots{
\Delta ^{im}={\cal W}(F)
}
The root system of the quiver is given by the real roots plus the imaginary 
roots. All real roots satisfy $\alpha \cdot C \cdot \alpha =2$ and all the
imaginary roots satisfy $\alpha \cdot C \cdot \alpha \leq 0$. 
We define the subset of the roots which can be written as
$\alpha=\sum_i k_i\alpha_i$ with $k_i\ge 0$
to be the {\it positive roots}.

For quivers without relations, Kac proved in \kac\ the following fundamental 
theorem: the dimension vectors of indecomposable representations are the 
positive roots of the quiver.  The real roots correspond to rigid
representations (i.e. with no moduli), while the imaginary roots
correspond to representations with moduli space dimension greater than
or equal to the expected dimension \quiverexpdim.  The possibility of
a greater actual dimension in this case is because there can be additional
unbroken gauge symmetry; conversely
for Schur roots (roots of Schur representations),
the true dimension will be equal to the expected dimension.

Finally, we quote a result useful in finding bound states.
For a quiver without relations, $\Ext^i(V,W)=0$ for $i\geq 2$, and
the Euler form gives the relative Euler character of two representations 
$V,W$ with dimension vectors $\alpha,\beta$:
\eqn\ekrhomext{
<\alpha,\beta>=\hbox {dim Hom}(\alpha,\beta)-\hbox {dim Ext}(\alpha, 
\beta)=\chi (\alpha,\beta)
}
which can be derived by considering the following exact sequence
\eqn\ekrexactsequ{
0\rightarrow \Hom (V,W)\rightarrow \sum _v \Hom (V_v,W_v)
\rightarrow \sum _a \Hom (V_{ta},W_{ha})\rightarrow \Ext (V,W)
\rightarrow 0
}

\subsec {An example: the generalized Kronecker quiver}

As an illustration of the definitions we have introduced in this appendix, 
we study in detail the following quiver 

\ifig\kronquiver{The Kronecker quiver}
{\epsfxsize2.0in\epsfbox{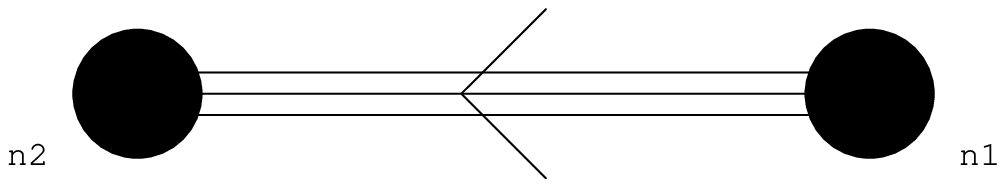}}

The Euler form for this quiver is given by
\eqn\ekrkronecker{
<\alpha ,\beta >=(\alpha _1\ \alpha _2)\left(\matrix{1&-3\cr 0&1}\right)
\left(\matrix{\beta _1\cr \beta _2}\right).}

Our goal is to find the Schur roots. The first and easiest step is to find
the indecomposable roots, which by Kac theorem are the positive roots of this
quiver. The fundamental roots are just $(0\ 1)$ and $(1\ 0)$. The respective
Weyl reflections are

\eqn\weylone{
r_1=\left(\matrix{-1&3\cr 0&1}\right),}

\eqn\weyltwo{
r_2=\left(\matrix{1&0\cr 3&-1}\right),}

The fundamental region consists of the vectors $v$ for which $r_i(v)\geq v$,
so it is the interior of the cone spanned by the vectors $u_1=(2\ 3)$ 
and $u_2=(3\ 2)$. The imaginary roots are those that can be brought to 
the fundamental region through Weyl reflections. They correspond to the 
interior of the cone spanned by the null vectors $v_{1,2}=(1,{3\pm\sqrt{5}
\over 2})$. Note that for imaginary roots the dimension of their moduli space 
is $d>0$.

Finally the real roots are obtained by the action of the Weyl group upon the
fundamental roots. They correspond to the integer solutions of the equation
$1=n_1^2+n_2^2-3n_1n_2$, which is just another way to say that they have 
$d=0$.  Explicitly, they are given by $(a_i\ a_{i+1})$ and $(a_{i+1}\ a_i)$ 
with
\eqn\orbseries{
a_{i+2}=3a_{i+1}-a_i,\; a_1=0\; \hbox{and}\; a_2=1.
}

We should also mention that the dimension vectors for which the expected
dimension formula predicts a negative dimension correspond to representations
which are direct sums of (real) roots, so they are not indecomposable. 

Weyl reflections appear in the discussion not just as symmetries of
the Cartan matrix; in fact there are explicit ``reflection
operations'' which take a representation and produce another representation
with the Weyl reflected dimension vector.

\ifig\kronecker{Schur roots of the Kronecker quiver}{\epsfxsize2.0in\epsfbox
{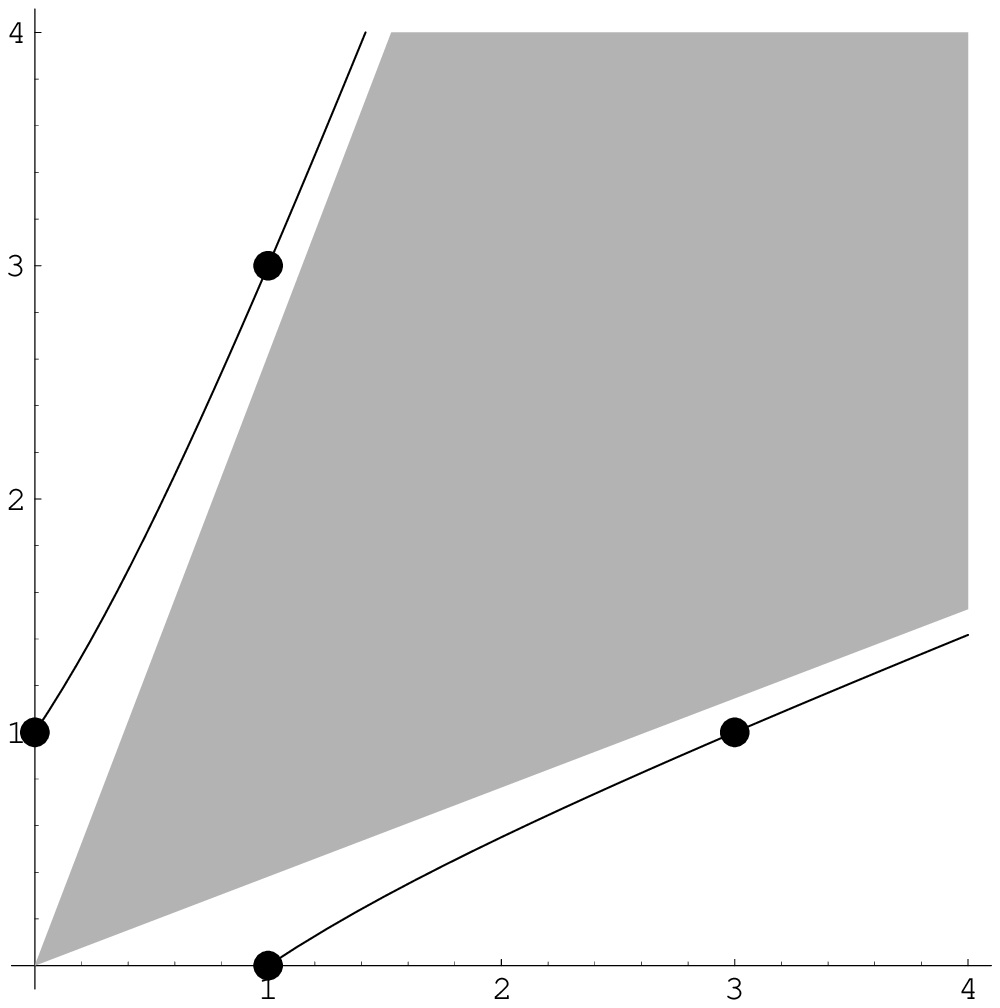}}

Having found the set of positive roots, by the Kac theorem we have the set of
dimension vectors for indecomposable representations. In general our next 
task will be to decide which of these correspond to Schur roots. 
The only characterization of Schur roots we are aware of is the following 
theorem due to Schofield, valid for quivers without relations \schofb

\eqn\schoth{\alpha\ \hbox { is Schur  }\Longleftrightarrow \ <\beta,\alpha>-
<\alpha,\beta>\ >\ 0 \hbox { for all } \beta \hbox { subvector of } \alpha}

In this simple example we can bypass this theorem, since
Kac \kac was able to prove that in this case all indecomposable roots are 
Schur (see also \schoa ). This is summarized in \kronecker.

After determining the Schur roots of this quiver, we can study their 
$\theta$-stability. Using Schofield's theorem \schoth, we learn that in
this case $(n_1\ n_2)$ is Schur if and only if $n_1'/n_2'<n_1/n_2$ for all 
its subvectors $(n_1'\ n_2')$. Introduce now a vector $(\theta _1\ 
\theta _2)$. A Schur root will be 
$\theta$-stable iff $n_1\theta _1+n_2\theta _2=0$ and $n_1'\theta _1+n_2'
\theta _2>0$. Using the characterization of Schur roots just given, this 
amounts to saying that a Schur root of the Kronecker quiver is $\theta$-stable 
iff $\theta _1<0$.

\subsec {Quivers with relations}

We consider a general set of relations, indexed by $r$, each expressed
as a weighted sum over paths starting at $ir$ and ending at $tr$.

Many of the basic definitions above generalize directly to this case.
In particular, there is an Euler form
\eqn\eulerrel{
<\alpha, \beta>=
\sum _v\alpha _v \beta _v-\sum _a \alpha _{ia}\beta _{ta}
+ \sum_r \alpha_{ir}\beta_{tr}
}
which stands in the same relation to the expected dimension 
as \eulercartan:
\eqn\expdimrel{
d = 1 + \sum_a \dim(ia) \dim(ta) - \sum_v \dim(v)^2 - 
\sum_{r} \dim(ir) \dim(tr) .
}
For the $\P^2$ Beilinson quiver, these forms are precisely the ones
\chiorbbasis\ and \cartanlv\ which appeared in the main text.

However, there do not seem to be results as general as the Kac theorem for
the general quiver with relations.  What results there are almost
all assume that the quiver contains no closed loops.
The specific
results we use for the $\P^2$ Beilinson quiver, such as the fact that
the expected dimension is realized, are generally taken from 
\refs{\lepot,\schob} and other works on bundles on $\P^2$.

This should probably not be too surprising as the
definition is general enough to encompass a wide variety of problems
in algebraic geometry and supersymmetric gauge theory.  
Of course we might still hope that 
the same structure of roots, the Weyl group
and so on will be useful in more general problems.
A basic point to check in this regard is whether
(in a given problem) the expected dimension is in fact
realized as an actual dimension of moduli space.
In general, as we discussed for the $\BZ_3$ quiver in section 4,
this need not be true, because of higher cohomology or equivalently relations
between relations.  

\appendix{B}{A few notations from algebraic geometry}


Some standard sheaves: $\CO_X$ is the sheaf of holomorphic functions
on $X$.  $\Omega$ is the holomorphic cotangent bundle.
$\Omega^n\equiv\Lambda^n\Omega$ is its $n$'th exterior
(antisymmetric) power.
$\Omega^d$ is the canonical line bundle (on a $d$-dimensional manifold).

On $\P^n$, $\CO(1)$ is the ``hyperplane bundle,''
whose sections have a zero on a single $\P^{n-1}\subset \P^n$.
(Thus they are linear combinations of the homogeneous coordinates).
$\CO(n)$ is the $n$-fold tensor product (sections are homogeneous
functions of degree $n$).  $E(n)$ for a general sheaf $E$ is
$E \otimes \CO(n)$.

For a bundle $E$, $H^n(M,E^*\otimes F)$ can be thought of as global 
holomorphic $n$-forms taking values in $E^* \otimes F$.  
The groups $\Ext^n(E,F)$ can be thought of as the sheaf (and more
generally homological algebra) generalization of this.  It will
generally satisfy the same ``topological'' properties, such as the
index theorem for $\chi=\sum (-1)^n\dim \Ext^n$, usually called the
Riemann-Roch theorem or Grothendieck-Riemann-Roch theorem.


Serre duality relates $H^n(M,E) \cong H^{d-n}(M,E^*\otimes\Omega^d)^*$.

\listrefs
\end